\documentclass[11pt,a4paper]{iopart}
\usepackage{bm}
\usepackage{hyperref}
\usepackage[square,numbers,sort&compress]{natbib}
\usepackage{graphicx}
\usepackage{amssymb}
\expandafter\let\csname equation*\endcsname\relax
\expandafter\let\csname endequation*\endcsname\relax
\usepackage{amsmath}
\usepackage{color}

\renewcommand{\d}{\mathrm{d}}

\renewcommand{\geq}{\geqslant}

\newcommand{\tauNL}{\tau_{\mathrm{NL}}}
\newcommand{\gNL}{g_{\mathrm{NL}}}
\newcommand{\fNL}{f_{\mathrm{NL}}}

\newcommand{\Sinit}{\mathcal{S}}
\newcommand{\Ainit}{\mathcal{A}}
\newcommand{\Binit}{\mathcal{B}}

\newcommand{\truncated}[1]{[\, \geq \text{{#1}p.f.} \,]}

\newcommand{\Mp}{M_{\mathrm{P}}}

\newcommand{\vect}[1]{\bm{\mathrm{{#1}}}}

\DeclareMathOperator{\pathorder}{\mathsf{P}}

\newcommand{\para}[1]{\par\vspace{2mm}\noindent\textbf{{#1}.---}}

\begin{document}

\title{Transport equations for the inflationary trispectrum}

\author{Gemma J. Anderson$^1$, David J. Mulryne$^2$ and David Seery$^1$}
\address{\vspace{2mm}
$^1$ Astronomy Centre, University of Sussex, Falmer Campus, Brighton, BN1 9QH,
UK\\[2mm]
$^2$ School of Physics and Astronomy, Queen Mary,
University of London, Mile End Road, London, E1 4NS, UK}

\eads{\mailto{G.Anderson@sussex.ac.uk}, \mailto{D.Mulryne@qmul.ac.uk},
\mailto{D.Seery@sussex.ac.uk}}

\begin{abstract}
	We use transport techniques to calculate the
	trispectrum produced in multiple-field inflationary models
	with canonical kinetic terms.
	Our method allows the time evolution of the local trispectrum parameters,
	$\tauNL$ and $\gNL$, to be tracked throughout the inflationary
	phase. We illustrate our approach using examples.
	We give a simplified method to calculate the superhorizon
	part of the relation between
	field fluctuations on spatially flat hypersurfaces and the
	curvature perturbation on uniform density slices, $\zeta$,
	and
	obtain its third-order part
	for the first time.
	We clarify how the `backwards' formalism of Yokoyama~et~al.
	relates to our analysis and other recent work.
	We supply explicit formulae which enable each inflationary observable
	to be computed in any canonical model of interest, using a
	suitable first-order ODE solver.
\end{abstract}
	
\maketitle

\section{Introduction}

Cosmological inflation predicts the generation of a primordial
perturbation, $\zeta$, believed to have seeded
the temperature anisotropy of the cosmic microwave background (``CMB'')
and the galaxy density field.
This fluctuation is sensitive to the physics that created it,
and therefore
different
models of inflation typically generate
perturbations with distinct statistical properties.
These properties can be observed by measuring their correlation functions.
We expect this approach to provide the most important observational
constraints on an era of early-universe inflation.

What information
is encoded in these correlation functions?
The two-point function is nearly determined by the symmetries of the
background, rather than the choice of microphysics,
although useful information may be extracted from
its scale dependence.
The higher $n$-point functions are much less constrained, but
only the three- and four-point functions (the
``bispectrum'' and ``trispectrum'') are likely to be measured in
the near future.
Canonical single field inflation predicts a bi- and trispectrum which
will be undetectable by present-day or near-future experiments
\cite{Maldacena:2002vr,Lyth:2005fi,Seery:2005gb,Seery:2006vu,
Seery:2006js,Seery:2008ax}.
But if more than one field is active during inflation, or
noncanonical interactions are present, the three- and four-point functions
can be measured and their properties can discriminate between these
possibilities.

Because of their observational relevance and
constraining power,
these ``nongaussian'' effects have received considerable attention.
During inflation, each comoving $k$-mode
of a light scalar field receives a perturbation when the corresponding
physical scale crosses outside the horizon.
Once outside, causality forbids any exchange between neighbouring
regions and
therefore $\zeta$ must be
generated by reprocessing the local fluctuations.
Where only a single degree of freedom $\zeta_g$ is relevant, this gives
\cite{Starobinsky:1986fxa,Sasaki:1995aw,Lyth:2005fi}
\begin{equation}
	\zeta(\vect{x}) = \zeta_g(\vect{x}) +
	\frac{3}{5} \fNL ( \zeta_g^2(\vect{x}) - \langle \zeta_g^2 \rangle )
	+ \frac{9}{25} \gNL \zeta_g^3(\vect{x}) + \cdots ,
	\label{eq:zeta-taylor}
\end{equation}
where all quantities are evaluated at the same time, and
$\vect{x}$ labels a coarse-grained spatial position with
sub-horizon details smoothed out.
This local character gives each correlation function a very
distinctive momentum dependence.
At leading-order the bispectrum has only one possibility,
generated by
the quadratic term in~\eqref{eq:zeta-taylor}.
Its amplitude is parametrized by the number $\fNL$
\cite{Verde:1999ij,Komatsu:2001rj},
which may depend weakly on the smoothing scale.
But the trispectrum has two possibilities,
generated respectively by the cubic term and
the square of the quadratic term.
These are conventionally parametrized by
the numbers $\tauNL$ and $\gNL$
\cite{Sasaki:2006kq,Boubekeur:2005fj,Alabidi:2005qi,Seery:2006js,
Byrnes:2006vq}.
In the single-field case,
$\tauNL$ does not appear in~\eqref{eq:zeta-taylor}
and can be expressed in terms of $\fNL$;
the precise relation is
$\tauNL = (6\fNL/5)^2$.
Where more than one light degree of freedom is present, they may all appear in
Eq.~\eqref{eq:zeta-taylor}
and this relation is weakened to the Suyama--Yamaguchi inequality
$\tauNL \geq (6\fNL/5)^2$ \cite{Suyama:2007bg,Smith:2011if}.
The role of such relations in diagnosing the active particle spectrum
during inflation was recently emphasized by
Assassi~et~al.~\cite{Assassi:2012zq}.

\para{Transport methods}%
In this paper we explain how the
non-linearity parameters $\tauNL$ and $\gNL$ can be calculated using
``transport'' methods.

Such calculations can already be carried out
within the ``$\delta N$
formalism''~\cite{Starobinsky:1986fxa,Sasaki:1995aw,Lyth:2005fi},
which requires a Taylor expansion of the background solution
in small displacements from a chosen initial condition.
An expression for $\tauNL$ was given in this formalism by
Alabidi \& Lyth~\cite{Alabidi:2005qi}.
A comparable result for $\gNL$
was provided
by Sasaki, V\"{a}liviita \& Wands \cite{Sasaki:2006kq}
in the context of a curvaton model,
and later generalized to an arbitrary number of light fields in
Ref.~\cite{Seery:2006js}.
The ``$\delta N$'' Taylor expansion leads to concise and
attractive analytic results.
But it is not ideally suited to numerical implementation,
because it relies on extracting small variations which can
easily be swamped by numerical noise.

In Ref.~\cite{Seery:2012vj} it was explained that the
Taylor expansion
can be understood as a variational method to
compute Jacobi fields for the flow of inflationary
trajectories in phase
space. These fields can be used to
explore local properties of any flow,
and were introduced by Jacobi in his reformulation of
Hamiltonian mechanics into what is
now Hamilton--Jacobi theory.
In inflation, they represent the geometrical structure which underlies
perturbation theory in the long wavelength limit.
They recur in many areas of physics
(see, e.g., Refs.~\cite{Hawking:1973uf,Visser:1992pz}),
and have been
much-studied in WKB approximations to the path integral
\cite{DeWittMorette:1976up,DeWittMorette:1984du,DeWittMorette:1984dw}.

The Jacobi fields are the necessary ingredient to compute
$\tauNL$ and $\gNL$, but
it is not necessary to use variational techniques
to compute them.
Their evolution can be determined
equally well using an
ordinary differential equation---the `Jacobi
equation'~\cite{Jacobi}.
The equivalence was emphasized by
DeWitt--Morette \cite{DeWittMorette:1976up}.
The Jacobi equation is usually preferable for numerical implementation.
It can be solved using conventional ODE techniques
and is usually much more stable against numerical noise.
Jacobi methods are widely used in other applications,
including gravitational
lensing~\cite{Lewis:2006fu,Lewis:2012tc}.

With this motivation, one can ask whether it
is possible to replace the ``$\delta N$'' Taylor expansion with an
approach based on the Jacobi equation.
To do so, one gives an evolution equation for each
$n$-point function.
Such equations were introduced
in Refs.~\cite{Mulryne:2009kh,Mulryne:2010rp}
and were originally framed in
real space.%
	\footnote{A similar formalism had been introduced earlier by
	Yokoyama, Suyama \& Tanaka \cite{Yokoyama:2007uu,Yokoyama:2007dw},
	who gave evolution equations for the Taylor coefficients
	of the $\delta N$ formalism rather than the $n$-point functions
	directly.
	It was shown in Ref.~\cite{Seery:2012vj} that these formalisms
	are equivalent up to the 3-point function.
	In \S\S\ref{sec:taylor}--\ref{sec:backwards} we extend this
	equivalence to the 4-point function.}
Real-space methods are adequate if one wishes to extract only
the local part of the three-point function.
But if one wishes to include more general momentum-dependence
or study $n$-point functions for $n \geq 4$, where it is necessary
to distinguish between
``squeezed'' and ``collapsed'' configurations,
one must revert to Fourier space.
In Ref.~\cite{Seery:2012vj} it was explained how to formulate
evolution equations for the full $\vect{k}$-space correlation functions,
which can be integrated using an approach similar
to the ``line of sight integral'' used to simplify
solution of the Boltzmann equation.
In Ref.~\cite{Seery:2012vj} this was used to give formal but explicit
expressions for the $n$-point functions in terms of the Jacobi fields
and their derivatives, and hence to demonstrate equivalence with the
variational ``$\delta N$ formalism'' up the three-point function.

In this paper we
specialize this method to the trispectrum.
We write a transport equation for the four-point function
of field fluctuations $\delta \varphi_\alpha$ defined on spatially flat
slices.
As in Ref.~\cite{Seery:2012vj}, this can be integrated in terms
of Jacobi fields
and reproduces the variational formulae discussed above.
In a second step, we
express the correlation functions of
$\zeta$ in terms of those of the $\delta \varphi_\alpha$.
At this point
the required values of $\tauNL$ and $\gNL$ can be extracted.
However, our method is not limited to obtention of
the $\zeta$ correlation functions
and can be deployed to determine the correlation functions
of both $\zeta$ and any isocurvature modes.

\para{Outline}%
In \S\ref{sec:transport} we introduce the transport framework and
extend it to third order.
In \S\ref{sec:evolution}
we write down the full $\vect{k}$-dependent equation
which evolves the four-point function on superhorizon scales.
By studying the momentum-dependence of this equation, we can extract
(in~\S\ref{sec:local-separation})
the coefficients of the ``squeezed''
and ``collapsed'' configurations.
We give separate evolution equations for these.

In~\S\ref{sec:taylor} we demonstrate that the transport (Jacobi)
method is equivalent to the familiar Taylor expansion of the
separate universe formalism.
We use our evolution equation to derive ordinary differential
equations which evolve the separate-universe Taylor coefficients
\emph{forward} in time,
and which supply the basis of an efficient numerical implementation.
In~\S\ref{sec:gauge-transform} we finish the task of
extracting $\tauNL$ and $\gNL$ by computing the relationship between
$\zeta$ and the field fluctuations $\delta\varphi_\alpha$.
Our final expressions are given in~\S\ref{sec:tau-g-formulae}.
We supply explicit expressions which enable each inflationary
observable to be computed in any canonical model of interest,
using a suitable first-order ODE solver.

In~\S\ref{sec:backwards} we describe the alternative
\emph{backwards} transport method introduced by Yokoyama~et~al.,
and extend it to accommodate the trispectrum parameters.
We briefly comment on the relative advantages of each formulation.
In~\S\ref{sec:results} we discuss some representative
numerical results. Finally, we conclude with a short discussion
in~\S\ref{sec:conclusions}.

\para{Notation and conventions}%
We set $c = \hbar = 1$ and work in terms of the reduced Planck mass,
$\Mp^2 = (8\pi G)^{-1}$ where $G$ is Newton's constant.
The species of light scalar fields are indexed by Greek labels
$\alpha$, $\beta$, \dots, .

\section{Transport Equations}
\label{sec:transport}

After smoothing on a length scale $k^{-1}\gg (aH)^{-1}$,
the field value in each smoothed region of the universe (``patch'')
will evolve independently,
as though it were in a homogeneous and isotropic
\emph{separate universe}.
Making use of the slow-roll approximation,
and
assuming that all fields are canonically normalized
and minimally coupled to Einstein gravity,
each smoothed field $\varphi_\alpha$ evolves according to
\cite{Mulryne:2009kh,Mulryne:2010rp,Seery:2012vj}
\begin{equation}
	\label{eq:u}
	\frac{\d\varphi_\alpha}{\d N} =
		- \Mp^2 \frac{\partial \ln V(\varphi)}
			{\partial \varphi_\alpha} \equiv u_{\alpha} ,
\end{equation}
up to gradient-suppressed corrections.
In writing~\eqref{eq:u}
we have used the e-folding number $\d N = H \, \d t$
as a time variable, and $t$ is cosmic time.
The index $\alpha$ labels the species of light scalar fields
and
$u_{\alpha}$ can be interpreted as a flow vector describing the trajectory
of the smoothed field in phase space.
In this paper we
will take these indices to be contracted using the
flat
metric $\delta_{\alpha\beta}$, so that index placement is
immaterial.

If desired the slow-roll approximation could be abandoned
by passing to a Hamiltonian formulation.
The resulting transport equations are structurally identical, requiring only
specification of suitable
initial conditions.
This method was described in Refs.~\cite{Mulryne:2010rp,Seery:2012vj}
and later implemented by Dias, Frazer \& Liddle \cite{Dias:2012nf}
for the purpose of studying D-brane models of inflation.
In this paper we will restrict ourselves to the slow-roll approximation,
but our evolution equations are unchanged by this choice and can
be extended immediately to the full phase space.

\subsection{Jacobi equation}
The field value varies between coarse-grained patches.
Picking a fiducial patch labelled by the spatial position
$\vect{x}$, the field in a neighbouring patch at
relative position $\vect{r}$
will be displaced by a small amount
$\delta \varphi_\alpha$,
\begin{equation}
	\varphi_{\alpha}(\vect{x} + \vect{r})
	\approx
	\varphi_{\alpha}(\vect{x}) + \delta\varphi_{\alpha}(\vect{r}) .
\end{equation}
At a generic position, and
provided the region under consideration is not too large,
we can expect $|\delta\varphi_{\alpha}|$ to be small in comparison with
$|\varphi_\alpha|$.
With these assumptions
the evolution of $\delta\varphi_{\alpha}$
can be obtained by making a Taylor expansion of the velocity
$u_{\alpha}$ in the neighbourhood of the fiducial trajectory.
Hence,
\begin{equation}
	\begin{split}
	\frac{\d \delta \varphi_{\alpha}(\vect{r})}{\d N}
		= \mbox{}
		&
		u_{\alpha\beta}[\varphi(\vect{x})]
		\delta \varphi_\beta(\vect{r})
		+
		\frac{1}{2!} u_{\alpha\beta\gamma}[\varphi(\vect{x})]
		\delta \varphi_\beta(\vect{r}) \delta \varphi_\gamma(\vect{r})
		\\ & \mbox{}
		+
		\frac{1}{3!} u_{\alpha\beta\gamma\delta}[\varphi(\vect{x})]
		\delta \varphi_\beta(\vect{r}) \delta \varphi_\gamma(\vect{r})
			\delta \varphi_\delta(\vect{r})
		+
		\cdots .
	\end{split}
	\label{eq:deviation}
\end{equation}
We now exchange $\vect{r}$ for a Fourier space description.
To keep the resulting equations compact we employ the `primed'
DeWitt index convention
introduced in Ref.~\cite{Seery:2012vj}.
In this notation, a compound index
such as $\alpha'$ includes a field label $\alpha$ and a momentum label
$\vect{k}_\alpha$, and also indicates evaluation at some common time
of interest $N$.
The summation convention applied to $\alpha'$
implies integration over momentum with measure
$\d^3 \vect{k}_\alpha / (2\pi)^3$, and
summation over the species $\alpha$.
In this notation we find
\begin{equation}
	\label{eq:jacobi}
	\begin{split}
		\frac{\d \delta \varphi_{\alpha'}}{\d N}
		= \mbox{}
		&
			u_{\alpha' \beta'} \delta \varphi_{\beta'}
			+
			\frac{1}{2!}
			u_{\alpha' \beta' \gamma'}
			\Big(
				\delta \varphi_{\beta'} \delta \varphi_{\gamma'}
				-
				\langle
					\delta \varphi_{\beta'} \delta \varphi_{\gamma'}
				\rangle
			\Big)
		\\ & \mbox{}
			+
			\frac{1}{3!}
			u_{\alpha' \beta' \gamma' \delta'}
			\Big(
				\delta \varphi_{\beta'} \delta \varphi_{\gamma'}
				\delta \varphi_{\delta'}
				-
				\langle
					\delta \varphi_{\beta'} \delta \varphi_{\gamma'}
					\delta \varphi_{\delta'}
				\rangle
			\Big)
			+
			\cdots
			.
	\end{split}
\end{equation}
Eq.~\eqref{eq:jacobi} is the nonlinear Jacobi equation.
We have subtracted a zero-mode, which amounts to discarding disconnected
terms in the correlation functions.
The $u$-matrices contained in~\eqref{eq:jacobi} inherit a dependence
on the fiducial region $\vect{x}$ through their dependence on the
background fields, but
the resulting connected correlation functions
depend only on statistical properties of the ensemble of smoothed fields.
Explicitly, we find
\begin{align}
	\label{eq: u2}
	u_{\alpha^{\prime}\beta^{\prime}} &
	\equiv
		(2\pi)^3
		\delta(\vect{k}_{\alpha} - \vect{k}_{\beta})
		u_{\alpha\beta}[\varphi(\vect{x})] \\
	\label{eq: u3}
	u_{\alpha^{\prime}\beta^{\prime}\gamma^{\prime}} &
	\equiv
		(2\pi)^{3}
		\delta(\vect{k}_{\alpha} - \vect{k}_{\beta} - \vect{k}_{\gamma})
		u_{\alpha\beta\gamma}[\varphi(\vect{x})] \\
	\label{eq: u4}
	u_{\alpha^{\prime}\beta^{\prime}\gamma^{\prime}\delta^{\prime}} &
	\equiv
		(2\pi)^{3}
		\delta(\vect{k}_{\alpha} - \vect{k}_{\beta} - \vect{k}_{\gamma}
		- \vect{k}_{\delta})
		u_{\alpha\beta\gamma\delta}[\varphi(\vect{x})] .
\end{align}

\subsection{Evolution of correlation functions}
\label{sec:evolution}

The Jacobi equation~\eqref{eq:jacobi}
summarizes evolution in the ensemble of smoothed patches.
The $u$-matrices can be calculated using any suitable method,
such as the long-wavelength limit of
cosmological perturbation theory or the separate-universe approximation.
However they are obtained,
they control not only the evolution of
physical field fluctuations
but also their correlation functions.

To show this we note that
for any classical observable $O$
not explicitly depending on time,
the time derivative of its expectation value satisfies
$\d \langle O \rangle / \d N =
\langle \d O / \d N \rangle$, provided probability is conserved.%
	\footnote{Technically, the probability
	distribution $P$ must vanish sufficiently rapidly
	on the boundary of phase space that $u_\alpha P \rightarrow 0$
	there, and therefore integration by parts inside the
	expectation value does not generate
	any boundary terms.}
It also applies quantum-mechanically if $O$ is a Heisenberg
picture field.
Transport equations for the quantum case, similar to those we
will develop here, were given by
Andrews \& Hall~\cite{Andrews:1985wn}
and developed by Ballentine \& McRae~\cite{Ballentine:1998wr}.
The classical limit was studied by Hepp~\cite{Hepp:1974br}.

We define the two-point function $\Sigma_{\alpha' \beta'}$ to satisfy
\begin{equation}
	\label{eq:ehrenfest}
	\Sigma_{\alpha^{\prime}\beta^{\prime}}
	\equiv
		\langle
			\delta \varphi_{\alpha^{\prime}} \delta \varphi_{\beta^{\prime}}
		\rangle .
\end{equation}
Recall that our index convention implies that each quantity on the
right-hand side is evaluated at the common time of interest, $N$.
Differentiating this expression, and moving the time derivative
inside the expectation value as discussed above,
we obtain an evolution equation
for $\Sigma_{\alpha' \beta'}$,
\begin{equation}
	\frac{\d\Sigma_{\alpha^{\prime}\beta^{\prime}}}{\d N}
	=
		\left\langle
			\frac{\d\delta\varphi_{\alpha^{\prime}}}{\d N}
			\delta \varphi_{\beta^{\prime}}
			+
			\delta \varphi_{\alpha^{\prime}}
			\frac{\d\delta\varphi_{\beta^{\prime}}}{\d N}
		\right\rangle .
\end{equation}
Use of Eq.~\eqref{eq:jacobi} allows the right-hand side to be rewritten in
terms of $u$-matrices and correlation functions.
Working to the lowest relevant order,%
	\footnote{Retaining higher-order contributions would reproduce the
	`loop corrections' of the $\delta N$ formalism; see
	Refs.~\cite{Boubekeur:2005fj,Lyth:2006qz,Seery:2010kh}.}
we conclude
\begin{equation}
	\label{eq: 2pf te}
	\frac{\d\Sigma_{\alpha^{\prime}\beta^{\prime}}}{\d N}
	=
		u_{\alpha^{\prime}\gamma^{\prime}}
		\Sigma_{\gamma^{\prime}\beta^{\prime}}
		+
		u_{\beta^{\prime}\gamma^{\prime}}
		\Sigma_{\gamma^{\prime}\alpha^{\prime}}
		+
		\truncated{3} ,
\end{equation}
where ``$\truncated{3}$'' denotes
terms containing
higher-order correlation functions which
have been omitted, beginning with the three-point function.
Eq.~\eqref{eq: 2pf te} will be a good approximation whenever these
higher-order correlation functions are negligible, which will usually
be satisfied during an epoch of quasi-exponential inflation.
In that case, the correlation functions typically order their
amplitudes in powers of $H^2$ \cite{Jarnhus:2007ia}
making the relative error after translation to $\zeta$ of order
$(H/\Mp)^2 \ll 1$.
A similar procedure gives the evolution of the three-point function.
We define
\begin{equation}
	\alpha_{\alpha^{\prime}\beta^{\prime}\gamma^{\prime}}
	\equiv
		\langle
			\delta \varphi_{\alpha^{\prime}} \delta \varphi_{\beta^{\prime}}
			\delta \varphi_{\gamma^{\prime}}
		\rangle ,
\end{equation}
and the corresponding transport equation is
\begin{equation}
	\label{eq: 3pf te}
	\frac{\d\alpha_{\alpha^{\prime}\beta^{\prime}\gamma^{\prime}}}{\d N}
	=
		u_{\alpha^{\prime}\lambda^{\prime}}
		\alpha_{\lambda^{\prime}\beta^{\prime}\gamma^{\prime}}
		+
		u_{\alpha^{\prime}\lambda^{\prime}\mu^{\prime}}
		\Sigma_{\lambda^{\prime}\beta^{\prime}}
		\Sigma_{\mu^{\prime}\gamma^{\prime}}
		+
		\text{cyclic}
		+ \truncated{4} ,
\end{equation}
where ``cyclic'' denotes the two cyclic permutations of each term,
and ``$\truncated{4}$'' again denotes terms involving higher-order
correlation functions which have been discarded, beginning with the
four-point function.
As for the two-point function, Eq.~\eqref{eq: 3pf te} will be a
good approximation whenever these are negligible in comparison with the
terms which have been retained.

\para{Four-point function}%
Eqs.~\eqref{eq: 2pf te} and~\eqref{eq: 3pf te}
were given in Ref.~\cite{Seery:2012vj}.
In this section, for the first time, we give the corresponding
transport equation for the four-point function.
To do so, we must distinguish carefully between the connected and
disconnected contributions.
The disconnected contributions are always present, even in the case of purely
Gaussian statistics, and therefore provide no new information.
But if
the perturbations develop some \emph{intrinsic} nongaussianity during
their evolution,
this is encoded in the \emph{connected} part of the four-point function.
To obtain it we subtract the disconnected terms from the full
four-point function, and define
\begin{equation}
	\beta_{\alpha^{\prime}\beta^{\prime}\gamma^{\prime}\delta^{\prime}}
	\equiv
		\langle
			\delta \varphi_{\alpha^{\prime}}
			\delta \varphi_{\beta^{\prime}}
			\delta \varphi_{\gamma^{\prime}}
			\delta \varphi_{\delta^{\prime}}
		\rangle
		-
		\Sigma_{\alpha' \beta'} \Sigma_{\gamma' \delta'}
		-
		\Sigma_{\alpha' \gamma'} \Sigma_{\beta' \delta'}
		-
		\Sigma_{\alpha' \delta'} \Sigma_{\beta' \gamma'} .
\end{equation}
In statistical language, the four-point function
$\langle \delta \varphi_{\alpha'} \delta \varphi_{\beta'}
\delta \varphi_{\gamma'} \delta \varphi_{\delta'} \rangle$
is the moment, and the connected part
$\beta_{\alpha' \beta' \gamma' \delta'}$ is the cumulant.

The transport equation for $\beta_{\alpha' \beta' \gamma' \delta'}$ is
\begin{equation}
	\label{eq: 4pf te}
	\begin{split}
	\frac{\d\beta_{\alpha' \beta' \gamma' \delta'}}{\d N}
	= \mbox{} &
		\Big(
			u_{\alpha' \lambda'}
			\beta_{\lambda' \beta' \gamma' \delta'}
			+
			\text{3 cyclic}
		\Big)
		+
		\Big(
			u_{\alpha' \lambda' \mu'}
			\alpha_{\lambda' \beta' \gamma'} \Sigma_{\mu' \delta'}
			+
			\text{11 cyclic}
		\Big)
	\\ & \mbox{}
		+
		\Big(
			u_{\alpha' \lambda' \mu' \nu'}
			\Sigma_{\lambda' \beta'} \Sigma_{\mu'\gamma'} \Sigma_{\nu' \delta'}
			+
			\text{3 cyclic}
		\Big)
		+
		\truncated{5} .
	\end{split}
\end{equation}
It can be obtained by various methods, including the Gauss--Hermite
cumulant expansion used in Ref.~\cite{Mulryne:2009kh},
the method of generating functions used in Ref.~\cite{Mulryne:2010rp}
and the approach described above.
 
\subsection{Separation of local shapes}
\label{sec:local-separation}

The transport equations~\eqref{eq: 2pf te}, \eqref{eq: 3pf te}
and~\eqref{eq: 4pf te} evolve each correlation function in its
entirety.
Although they are first order ordinary differential equations,
they are not trivial to solve because they couple
the correlation functions associated with different $\vect{k}$- and
species labels.%
	\footnote{For example, the four-point function with
	momentum labels $\vect{k}_1$, $\vect{k}_2$, $\vect{k}_3$
	and $\vect{k}_4$ couples to other correlation functions
	with momenta $\vect{k}_1 + \vect{k}_2$, and so on.
	Had we retained loop corrections, these would
	make the hierarchy considerably more complex because
	each correlation function no longer couples only to a few other
	isolated $\vect{k}$-modes, but to the whole phase space of
	soft superhorizon modes. Handling this is a computational
	challenge. For one approach see, eg., Ref.~\cite{Huston:2011vt}.}
Indeed,
the coupled system can be regarded as simply a form of Boltzmann hierarchy.
Like the hierarchy used to compute CMB anisotropies it must be
truncated---by discarding higher-order
correlation functions---if
it is to be turned into a practical computational tool.
We will see in~\S\ref{sec:taylor} that it admits a similar
kind of formal solution.
But
if we wish only to track the evolution of the local
momentum shapes, then we can extract simpler ``flavour'' equations 
which do not involve the continuum of $\vect{k}$-modes.
These are ordinary differential equations for a finite number of variables
and their numerical solution is straightforward.

Eqs.~\eqref{eq: 2pf te}, \eqref{eq: 3pf te} and~\eqref{eq: 4pf te}
show that (at least to this order), each correlation function is
sourced
by the correlation functions of \emph{lower}
order. Hence, we proceed inductively: if the $\vect{k}$-dependence of the
two-point function is known, then it can be used to determine the
local $\vect{k}$-dependence inherited by the three-point function and
subsequently the four-point function.

\para{Two-point function}%
Since we anticipate approximate scale-invariance, we write the
two-point function as
\begin{equation}
	\Sigma_{\alpha^{\prime}\beta^{\prime}}
	=
		(2\pi)^3
		\delta(\vect{k}_\alpha + \vect{k}_\beta)
		\frac{\Sigma_{\alpha\beta}}{k_{\alpha}^{3}} ,
	\label{eq: 2pf def}
\end{equation}
where $\Sigma_{\alpha\beta}$ has dimension of $[\text{mass}]^2$
but is nearly independent of $k_\alpha = k_\beta$.
It is this $k^{-3}_\alpha$ dependence which will be inherited by
all higher $n$-point functions. The possible ways in which this inheritance
can happen correspond to the possible local
(``squeezed'' and ``collapsed'') momentum shapes.

We first require a transport equation for $\Sigma_{\alpha\beta}$.
As described above,
this is a flavour-only matrix, carrying indices for the species of
scalar fields but not momentum labels.
Substituting~\eqref{eq: 2pf def} into~\eqref{eq: 2pf te}, we conclude
\begin{equation}
	\label{eq: 2pf flavour}
	\frac{\d \Sigma_{\alpha\beta}}{\d N}
	=
		u_{\alpha \lambda} \Sigma_{\lambda \beta}
		+
		u_{\beta \lambda} \Sigma_{\lambda \alpha} .
\end{equation}
This is symbolically the same equation as the full $\vect{k}$-space
transport equation, Eq.~\eqref{eq: 2pf te}, with
primed indices exchanged for unprimed ones.

In practice, $\Sigma_{\alpha\beta}$ carries a small dependence on the
$k$-scale at which it is evaluated.
This $k$-dependence, typically characterized by a spectral index,
can also be calculated by transport methods;
see Dias~et~al.~\cite{Dias:2011xy}.
Recently Dias, Frazer \& Liddle extended this method to obtain the
scale-dependence of the spectral index, or ``running''
\cite{Dias:2012nf}.

\para{Three-point function}%
Examination of the transport equation for the three-point function,
Eq.~\eqref{eq: 3pf te}, shows that in a small time interval
$\delta N$, the change to $\alpha_{\alpha' \beta' \gamma'}$
is of the schematic form
$\delta \alpha \sim ( u' \alpha + u'' \Sigma \Sigma ) \delta N$,
where a prime $'$ applied to $u$ indicates one of the
field-space derivatives which generate the index structure for the
$u$-matrices.
The $u' \alpha$ terms generate a change $\delta \alpha$ which is
proportional to the momentum-dependence already carried by $\alpha$.
Therefore this term can reorganize the amplitudes of these shapes,
but introduces no new types of momentum dependence.
New shapes are sourced
only by the $\Sigma \Sigma$ terms.

Eq.~\eqref{eq: 2pf def} shows that the product $\Sigma \Sigma$
must generate a
shape of the form $k_{\alpha}^{-3} k_{\beta}^{-3}$,
and therefore
the most general structure which can be \emph{sourced} during the
evolution has the form
\begin{equation}
	\label{eq: 3pf def}
	\alpha_{\alpha^{\prime}\beta^{\prime}\gamma^{\prime}}
	\supseteq
		(2\pi)^3\delta(\vect{k}_\alpha + \vect{k}_\beta + \vect{k}_\gamma)
		\bigg(
			\frac{a_{\alpha\mid\beta\gamma}}{k_{\beta}^{3}k_{\gamma}^{3}}
			+
			\frac{a_{\beta\mid\alpha\gamma}}{ k_{\alpha}^{3}k_{\gamma}^{3}}
			+
			\frac{a_{\gamma\mid\alpha\beta}}{k_{\alpha}^{3}k_{\beta}^{3}}
		\bigg)
	,
\end{equation}
where we use
the notation ``$\supseteq$'' to indicate that the three-point function
contains this term together with others which have not been written.
The matrices $a_{\alpha\mid\beta\gamma}$ are symmetric under exchange
of $\beta \leftrightarrow \gamma$, but need not possess further symmetries.
The full three-point function corresponds to the sourced
contribution~\eqref{eq: 3pf def} plus an \emph{unsourced}
term appearing as its initial condition. The unsourced
piece is generated by quantum interference effects operating around the
epoch of horizon exit, and typically has a very complicated
momentum dependence \cite{Seery:2005gb}.
However, its amplitude is small
in the canonical models to which we
restrict attention in this paper
\cite{Lyth:2005qj,Vernizzi:2006ve}.

After substitution of~\eqref{eq: 3pf def} into the transport
equation~\eqref{eq: 3pf te}, we obtain an evolution equation for
$a_{\alpha\mid\beta\gamma}$,
\begin{equation}
	\label{eq: 3pf flavour}
	\frac{\d a_{\alpha\mid\beta\gamma}}{\d N}
	=
		u_{\alpha\lambda} a_{\lambda\mid\beta\gamma}
		+
		u_{\beta\lambda} a_{\alpha\mid\lambda\gamma}
		+
		u_{\gamma\lambda} a_{\alpha\mid\beta\lambda}
		+
		u_{\alpha\lambda\mu} \Sigma_{\lambda\beta} \Sigma_{\mu\gamma} .
\end{equation}
Eq.~\eqref{eq: 3pf flavour} strictly applies only when the
momenta entering the correlation function are not too dissimilar
in magnitude. This is usually an acceptable approximation for
CMB experiments, but a more refined analysis might be required where
larger hierarchies of scale exist. This issue is not confined to the
transport framework; it applies to results obtained using
any method, including the familiar $\delta N$
Taylor expansion.

\para{Four-point function}%
Eqs.~\eqref{eq: 2pf flavour} and~\eqref{eq: 3pf flavour} were given in
Ref.~\cite{Seery:2012vj}.
The same analysis applied to the four-point function shows that, in
a small time interval $\delta N$, the change in the
connected part of the correlation function
has the schematic form
\begin{equation}
	\delta \beta \sim
		( u' \beta + u'' \alpha \Sigma + u ''' \Sigma \Sigma \Sigma )
		\delta N .
\end{equation}
As for the three-point function, the term $u' \beta$ is simply a
shift in the amplitude of shapes already present in $\beta$.
The sourced contributions are now $u'' \alpha \Sigma$ and
$u''' \Sigma \Sigma \Sigma$. Of these, the $\Sigma \Sigma \Sigma$ term
must generate a shape of the form $k_{\alpha}^{-3} k_{\beta}^{-3}
k_{\gamma}^{-3}$, which can be recognized as a $\gNL$-type contribution
\cite{Byrnes:2006vq}.

The $\alpha \Sigma$ term is more complex, because the momentum
$\delta$-function in $u''$ [see Eq.~\eqref{eq: u3}]
reorganizes the momenta appearing
in the denominators of the three-point function~\eqref{eq: 3pf def}.
Written out explicitly, this term is
\begin{equation}
	\label{eq: 4pf explicit}
	\begin{split}
	u_{\alpha' \lambda' \mu'}
	\alpha_{\lambda' \beta' \gamma'}
	\Sigma_{\mu' \delta'}
	=
		(2\pi)^3 \!
		\int \!
		\d^3 k_{\lambda} \,
		\d^3 k_{\mu}
		\,
		&
		\delta(\vect{k}_\alpha - \vect{k}_\lambda - \vect{k}_\mu)
		\delta(\vect{k}_\lambda + \vect{k}_\beta + \vect{k}_\gamma)
		\delta(\vect{k}_\mu + \vect{k}_\delta)
		\\ & \mbox{} \times u_{\alpha \lambda \mu}
		\frac{\Sigma_{\mu\delta}}{k_\delta^3}
		\bigg(
			\frac{a_{\lambda\mid\beta\gamma}}{k_\beta^3 k_\gamma^3}
			+
			\frac{a_{\beta\mid\lambda\gamma}}{k_\lambda^3 k_\gamma^3}
			+
			\frac{a_{\gamma\mid\beta\lambda}}{k_\beta^3 k_\lambda^3}
		\bigg) ,
	\end{split}
\end{equation}
plus the nontrivial permutations of
$\alpha'$, $\beta'$, $\gamma'$ and $\delta'$.
The first term in round brackets, $\sim k_{\beta}^{-3} k_{\gamma}^{-3}$,
has the form of a $\gNL$-type contribution.
But the remaining terms involve $k_{\lambda}^{-3}$,
and the $\delta$-functions in~\eqref{eq: 4pf explicit}
show that $\vect{k}_\lambda = \vect{k}_\alpha + \vect{k}_\delta$.
Therefore this term generates a \emph{different} momentum shape;
it is the ``collapsed'' configuration, which corresponds to a $\tauNL$-type
contribution \cite{Byrnes:2006vq}.
It follows that the most general structure sourced by time evolution
can be written
\begin{equation}
	\label{eq: 4pf def}
	\beta_{\alpha^{\prime}\beta^{\prime}\gamma^{\prime}\delta^{\prime}}
	=
		(2\pi)^3
		\delta(\vect{k}_\alpha + \vect{k}_\beta + \vect{k}_\gamma
			+ \vect{k}_\delta)
		\Big(
			\frac{g_{\alpha\mid\beta\gamma\delta}}
				{k_\beta^3k_\gamma^3k_\delta^3}
			+ \text{3 cyclic}
			+ \frac{\tau_{\alpha\beta\mid\gamma\delta}}
				{k_\alpha^3k_\beta^3 |\vect{k}_\alpha + \vect{k}_\gamma|^3}
			+ \text{11 cyclic}
		\Big)
	,
\end{equation}
where the ``cyclic'' pieces refer to the cyclic permutations of the
preceding terms.
The matrix $g_{\alpha\mid\beta\gamma\delta}$ is symmetric
under any exchange of ${\beta, \gamma, \delta}$, but has no symmetries
under permutations involving $\alpha$.
The matrix $\tau_{\alpha\beta\mid\gamma\delta}$ is symmetric under
the simultaneous exchanges $\alpha \leftrightarrow \beta$
and $\gamma \leftrightarrow \delta$, giving 12 independent
elements.

Substitution of~\eqref{eq: 4pf def} into the transport
equation~\eqref{eq: 4pf te}
enables us to extract individual evolution equations for
$g_{\alpha\mid\beta\gamma\delta}$ and
$\tau_{\alpha\beta\mid\gamma\delta}$. They are
\begin{align}
	\nonumber
	\frac{\d g_{\alpha\mid\beta\gamma\delta}}{\d N}
	& =
		u_{\alpha\lambda} g_{\lambda\mid\beta\gamma\delta}
		+
		u_{\beta\lambda} g_{\alpha\mid\lambda\gamma\delta}
		+
		u_{\gamma\lambda} g_{\alpha\mid\beta\lambda\delta}
		+
		u_{\delta\lambda} g_{\alpha\mid\beta\gamma\lambda}
		\\
	& \quad	\mbox{}
		+
		u_{\alpha\lambda\mu} a_{\lambda\mid\beta\gamma} \Sigma_{\mu\delta}
		+
		u_{\alpha\lambda\mu} a_{\lambda\mid\beta\delta} \Sigma_{\mu\gamma}
		+
		u_{\alpha\lambda\mu} a_{\lambda\mid\gamma\delta} \Sigma_{\mu\beta}
		+
		u_{\alpha\lambda\mu\nu}
			\Sigma_{\lambda\beta}
			\Sigma_{\mu\gamma}
			\Sigma_{\nu\delta}
	\label{eq: g te}
	\\ \nonumber
	\frac{\d \tau_{\alpha\beta\mid\gamma\delta}}{\d N}
	& =
		u_{\alpha\lambda} \tau_{\lambda\beta\mid\gamma\delta}
		+
		u_{\beta\lambda} \tau_{\alpha\lambda\mid\gamma\delta}
		+
		u_{\gamma\lambda} \tau_{\alpha\beta\mid\lambda\delta}
		+
		u_{\delta\lambda} \tau_{\alpha\beta\mid\gamma\lambda}
		\\
	& \quad \mbox{}
		+
		u_{\gamma\lambda\mu} \Sigma_{\mu\alpha} a_{\delta\mid\lambda\beta}
		+
		u_{\delta\lambda\mu} \Sigma_{\mu\beta} a_{\gamma\mid\lambda\alpha}
		.
	\label{eq: tau te}
\end{align}
Note that the $a$-dependent source terms in the second line
of~\eqref{eq: tau te} preserve the symmetry under simultaneous
exchange of the index pairs $(\alpha, \beta)$ and
$(\gamma, \delta)$.
We have dropped the initial value of $\alpha_{\alpha' \beta' \gamma'}$,
even though it appears in~\eqref{eq: 4pf te} as a source term
and, as a matter of principle, could appear in $\beta_{\alpha' \beta'
\gamma' \delta'}$ with a non-negligible coefficient.
In~\ref{appendix:ic} we show that this will usually be an acceptable
approximation
in models with canonically normalized scalar fields;
the initial value of $\alpha_{\alpha' \beta' \gamma'}$ remains
negligible
provided $|r\fNL| \lesssim 1$ throughout the evolution
where $r < 1$ is the tensor-to-scalar ratio.
On the other hand, in non-canonical models where the
initial value need \emph{not} be negligible it is important to retain this
term \cite{RenauxPetel:2009sj}.
 
\section{Equivalence to Taylor expansion method}
\label{sec:taylor}

Eqs.~\eqref{eq: g te}--\eqref{eq: tau te} enable us to follow the evolution
of the sourced, local-mode contributions to the trispectrum.
As we will explain in
\S\ref{sec:gauge-transform}, after
changing variable to $\zeta$
they allow us to calculate the
observable quantities $\tauNL$ and $\gNL$.
However, they are quite different in appearance to the familiar
expressions of the ``$\delta N$ formalism'',%
	\footnote{Here and below, we use the term ``$\delta N$ formalism''
	to mean a Taylor expansion in the initial conditions, even if the
	quantity being expanded is not $N$.}
which take the form of a Taylor expansion in the initial
conditions \cite{Lyth:2005fi}.

The connexion between these methods was explored in Ref.~\cite{Seery:2012vj}.
By formally integrating the transport equations, in a similar way to the
``line of sight'' integral
used when solving the Boltzmann equation, it is possible to demonstrate
equality with the ``$\delta N$'' expressions.
In Ref.~\cite{Seery:2012vj} this analysis was given for the two- and
three-point functions. Here we extend it to include the four-point
function.

\para{Integrating factor}%
The ``line of sight integral'' naturally expresses
each correlation function in terms of the underlying Jacobi fields.
We briefly recapitulate the argument of Ref.~\cite{Seery:2012vj}.
Without loss of generality, we write the two-point function
in the form
\begin{equation}
	\label{sigmasol}
	\Sigma_{\alpha'\beta'} =
	\Gamma_{\alpha' i'}\Gamma_{\beta' j'} \Sigma_{i' j'} .
\end{equation}
A suitable choice
for $\Gamma_{\alpha' i'}$ means it will function as an integrating factor.
In writing~\eqref{sigmasol} we have introduced a new type of primed
Latin index ($i'$, $j'$, \ldots). This has the same interpretation
as the primed Greek indices: $i'$ carries a flavour index $i$
and a momentum label $\vect{k}_i$, which range over the same values
as $\alpha$ and $\vect{k}_\alpha$. However, it indicates evaluation
at a different time $N_0$, as follows.
Substitution of Eq.~\eqref{sigmasol}
in~\eqref{eq: 2pf te} shows that the terms involving
$u_{\alpha'\beta'}$ can be removed if $\Gamma$ is chosen
to satisfy
\begin{equation}
	\label{eq:jacobi-equation}
	\frac{\d \Gamma_{\alpha' i'}}{\d N}
	=
		u_{\alpha' \beta'}\Gamma_{\beta' i'} .
\end{equation}
Comparison with Eqs~\eqref{eq:deviation}--\eqref{eq:jacobi}
shows that $\Gamma_{\alpha' i'}$ has an interpretation as
a differential coefficient,
\begin{equation}
	\Gamma_{\alpha' i'} =
		\frac{\partial \varphi_{\alpha'}}{\partial \varphi_{i'}}
		=
		\frac{\partial \varphi_\alpha (\vect{k}_\alpha, N)}
			{\partial \varphi_i(\vect{k}_i, N_0)}
		=
		\delta(\vect{k}_\alpha - \vect{k}_i)
		\frac{\partial \varphi_\alpha (N)}{\partial \varphi_i(N_0)}
	\label{eq:gamma-diff}
\end{equation}

Eq. \eqref{eq:gamma-diff}
is sometimes described
as the ``Jacobi map''.
It has a formal solution in terms of a path-ordered
exponential
\begin{equation}
	\label{eq:gamma-one}
	\Gamma_{\alpha' i'} = (2\pi)^3 \delta(\vect{k}_\alpha - \vect{k}_i)
	\pathorder \exp \left(
		\int_{N_0}^N \d N' \; u_{\alpha i}(N')
	\right) .
\end{equation}
In this expression, $\pathorder$ denotes the path-ordering operator
which rewrites its argument in order of position on the trajectory:
objects evaluated early on the trajectory appear to the right of
objects evaluated later.
This path-ordered exponential is related to the inverse of
the van Vleck
matrix, which is equivalent to the matrix of Jacobi fields.
Reference to Eqs.~\eqref{eq: 3pf te} and~\eqref{eq: 4pf te}
shows that, in each transport equation,
this choice for $\Gamma$
will absorb
the terms proportional to the $n$-point function itself.
Returning to the two-point function and
discarding higher-order contributions,
it follows that
the ``kernel'' $\Sigma_{i' j'}$ can be obtained as an integral over
the source terms.
It is this integral over sources which can be compared to the ``line of sight''
integral for the Boltzmann equation.

With these choices,
and working to leading order,
there are \emph{no} sources for the kernel
$\Sigma_{i' j'}$. Therefore it is constant, and equal to its initial
condition set at horizon crossing.
We write this constant value $\Sinit_{i' j'}$.

\para{Three-point function}%
When this method is applied
to the three-point function, it transpires that
the kernel \emph{is} sourced. Again without loss of generality, we write
\begin{equation}
	\label{ansatz:alpha}
	\alpha_{\alpha'\beta'\gamma'}
	=
		\Gamma_{\alpha' i'}
		\Gamma_{\beta' j'}
		\Gamma_{\gamma' k'}
		A_{i'j'k'} .
\end{equation}
We define $\tilde{u}_{i' j' k'} \equiv \Gamma^{-1}_{i' \alpha'}
u_{\alpha' \beta' \gamma'} \Gamma_{\beta' j'}\Gamma_{\gamma' k'}$
and obtain
\begin{equation}
	\label{eq:alpha-source-integral}
	A_{i' j' k'}
	=
		\Ainit_{i' j' k'}
		+
		\left[
			\int^N_{N_0} \d N' \;
			\tilde{u}_{i' m' n'}(N') \Sinit_{m' j'} \Sinit_{n' k'}
			+
			\text{2 cyclic}
		\right]
		+
		\Or( H^6 ) .
\end{equation}
The integration constant $\Ainit_{i' j' k'}$ is the unsourced initial
condition which was neglected above,
and the estimate $\Or(H^6)$ for the terms we have omitted
assumes that
the correlation functions order themselves in increasing powers of
$H^2$ as described by Jarnhus \& Sloth \cite{Jarnhus:2007ia}.
Defining
\begin{equation}
	\label{eq:gamma-two}
	\begin{split}
	\Gamma_{\alpha' i' j'}
	=
		\Gamma_{\alpha' m'}
		\int^N_{N_0} \tilde{u}_{m' i' j'}(N') \;  \d N'
	& =
		(2\pi)^3 \delta(\vect{k}_\alpha - \vect{k}_i - \vect{k}_j)
		\Gamma_{\alpha m}
		\int^N_{N_0} \tilde{u}_{mij}(N') \; \d N' ,
	\\ & =
		(2\pi)^3 \delta(\vect{k}_\alpha - \vect{k}_i - \vect{k}_j)
		\Gamma_{\alpha i j} ,
	\end{split}
\end{equation}
we conclude
\begin{equation}
	\label{eq:alpha-expansion}
	\alpha_{\alpha'\beta'\gamma'}
	=
		\Gamma_{\alpha' i'}\Gamma_{\beta' j'}\Gamma_{\gamma' k'}
		\Ainit_{i' j' k'}
		+
		\Big(
			\Gamma_{\alpha' i' j'} \Gamma_{\beta' k'} \Gamma_{\gamma' \ell'}
			\Sinit_{i' k'} \Sinit_{j' \ell'}
			+
			\text{2 cyclic}
		\Big)
		+
		\text{loops} .
\end{equation}
It can be shown that the quantity $\Gamma_{\alpha i j}$
appearing on
the right-hand side of~\eqref{eq:gamma-two}
is equal to%
	\footnote{Direct differentiation of Eq.~\eqref{eq:gamma-one} is
	subtle, because of the path-ordered exponential.
	It is simpler to differentiate
	the Jacobi equation, Eq.~\eqref{eq:jacobi-equation}, and then solve
	it using $\Gamma_{\alpha i}$ as an integrating factor.}
\begin{equation}
	\Gamma_{\alpha ij}
		=
		\frac{\partial \Gamma_{\alpha i}}{\partial \varphi_j(N_0)}
		=
		\frac{\partial^2 \phi_\alpha(N)}{\partial \varphi_i(N_0) 
										 \partial \varphi_j(N_0)} ,
\end{equation}
from which it
follows that Eq.~\eqref{eq:alpha-expansion}
is equivalent to the
Lyth--Rodr\'{\i}guez Taylor expansion formula for the
three-point function \cite{Lyth:2005fi}.
Moreover, differentiation of~\eqref{eq:gamma-two}
shows that $\Gamma_{\alpha ij}$
satisfies the evolution equation
\begin{equation}
	\label{eq:gamma2-evolve}
	\frac{\d \Gamma_{\alpha i j}}{\d N}
	= u_{\alpha \beta} \Gamma_{\beta i j} +
	u_{\alpha \beta \gamma} \Gamma_{\beta i} \Gamma_{\gamma j} .
\end{equation}

\para{Four-point function}%
The analysis for the four-point function is similar. We introduce
the integrating factor $\Gamma_{\alpha' i'}$,
\begin{equation}
	\beta_{\alpha' \beta' \gamma' \delta'} 
	=
		\Gamma_{\alpha' i'}
		\Gamma_{\beta' j'}
		\Gamma_{\gamma' k'}
		\Gamma_{\delta' \ell'}
		B_{i' j' k' \ell'} .
\end{equation}
The kernel $B_{i' j' k' \ell'}$ is given by an integral over sources,
as before, which are drawn from the lower-order $n$-point functions.
In this case they are the two- and three-point functions.
Keeping only leading-order terms, we find
\begin{equation}
	\label{eq:beta-source-integral}
	\begin{split}
	B_{i' j' k' \ell'}
	=
		\Binit_{i' j' k' \ell'}
		& \mbox{}
		+
		\Big(
			\int_{N_0}^N \d N' \;
				\tilde{u}_{i' p' q'}(N')
				A_{p' j' k'}(N')
				\Sinit_{q' \ell'}
			+
			\text{11 cyclic}
		\Big)
	\\ & \mbox{}
		+
		\Big(
			\int_{N_0}^N \d N' \;
				\tilde{u}_{i' q' r' s'}(N')
				\Sinit_{q' j'}
				\Sinit_{r' k'}
				\Sinit_{s' \ell'}
			+
			\text{3 cyclic}
		\Big)
		+
		\Or(H^8) ,
	\end{split}
\end{equation}
where we have defined
$\tilde{u}_{i' j' k' \ell'} = \Gamma^{-1}_{i' \alpha'}
u_{\alpha' \beta' \gamma' \delta'}
\Gamma_{\beta' j'} \Gamma_{\gamma' k'} \Gamma_{\delta' \ell'}$.
The integration constant $\Binit_{i' j' k' \ell'}$
is the initial value of the four-point function
at time $N = N_0$, as for the
three-point function.
Taking the four $\vect{k}$-modes entering
the four-point function to have a similar time of horizon
exit, and the initial time $N_0$ to be around this epoch,
the initial condition
was shown in Refs.~\cite{Seery:2006vu,Seery:2008ax}
to be dominated by the correlations induced by
decay of gravitational waves into scalar quanta.
It is negligible
when the amplitude of the four-point function is sufficiently large
to be observable.
On the other hand, the initial value of the three-point function
appears in the kernel $A_{p' j' k'}$ which forms part of the source
integral~\eqref{eq:beta-source-integral},
and need not be entirely negligible.
However,
as discussed below Eqs.~\eqref{eq: g te}--\eqref{eq: tau te},
and in more detail in~\ref{appendix:ic},
its contribution to $\tauNL$ or $\gNL$ is likely no more than
$\Or(1)$ for models with acceptable $|\fNL|$.

To relate~\eqref{eq:beta-source-integral} to the expressions
produced by the Taylor expansion algorithm we must express $A_{p' j' k'}$
purely in terms of correlations at the initial time $N_0$.
Combining~\eqref{eq:alpha-source-integral}
and~\eqref{eq:beta-source-integral}
we find
\begin{equation}
	\label{eq:beta-source-unpacked}
	\begin{split}
	\int_{N_0}^N \d N' \; & \tilde{u}_{i' p' q'}(N') A_{p' j' k'}(N')
	\\ &
	=
		\int_{N_0}^N \d N' \; \tilde{u}_{i' p' q'}(N')
		\Big\{
			\Ainit_{p' j' k'}
			+
			\int_{N_0}^{N'} \d N'' \;
				\tilde{u}_{p' r' s'}(N'') \Sinit_{r' j'} \Sinit_{s' k'}
		\\
		& \qquad \mbox{}
			+
			\int_{N_0}^{N'} \d N'' \;
				\tilde{u}_{j' r' s'}(N'') \Sinit_{r' p'} \Sinit_{s' k'}
			+
			\int_{N_0}^{N'} \d N'' \;
				\tilde{u}_{k' r' s'}(N'') \Sinit_{r' p'} \Sinit_{s' j'}
		\Big\}
	\end{split}
\end{equation}
The term involving $\Ainit_{p' j' k'}$ presents no difficulties.
It makes a contribution to $\beta_{\alpha' \beta' \gamma' \delta'}$
of the form
\begin{equation}
	\label{eq:A-P-term}
	\beta_{\alpha' \beta' \gamma' \delta'}
	\supseteq
		\Gamma_{\alpha' p' q'}
		\Gamma_{\beta' j'}
		\Gamma_{\gamma' k'}
		\Gamma_{\delta' \ell'}
		\Ainit_{p' j' k'}
		\Sinit_{q' \ell'}
		+ \text{11 cyclic} ,
\end{equation}
where, as above, the symbol ``$\supseteq$'' indicates that the four-point
function contains this contribution among others.
The other terms in~\eqref{eq:beta-source-unpacked}
are nested integrals, and divide into two groups.
One involves a contraction between the two $u$-matrices,
of the form
$\tilde{u}_{i' p' q'} \tilde{u}_{p' r' s'}$.
We first focus on the other two, which involve no contraction.
After summing over perturbations there are twenty-four such terms.
Consider the specific choice
$\tilde{u}_{i' p' q'} \tilde{u}_{j' r' s'}$
which appears in~\eqref{eq:beta-source-unpacked}.
In combination with one of the terms generated by
simultaneously exchanging 
$i' \leftrightarrow j'$ and $k' \leftrightarrow \ell'$
this generates
\begin{equation}
	\int_{N_0}^N \d N' \int_{N_0}^N \d N''
	\;
	\tilde{u}_{i' p' q'}(N')
	\tilde{u}_{j' r' s'}(N'')
	\Sinit_{r' p'}
	\Sinit_{s' k'}
	\Sinit_{q' \ell'} ,
	\label{eq:beta-source-product}
\end{equation}
in which the integrals are no longer nested.
Pairing all such terms in this way generates the 12 cyclic permutations
of indices in~\eqref{eq:beta-source-product}.
The corresponding contribution to the four-point function is
\begin{equation}
	\label{eq:tau-NL-term}
	\beta_{\alpha' \beta' \gamma' \delta'}
	\supseteq
		\Gamma_{\alpha' p' q'}
		\Gamma_{\beta' r' s'}
		\Gamma_{\gamma' k'}
		\Gamma_{\delta' \ell'}
		\Sinit_{p' r'}
		\Sinit_{s' k'}
		\Sinit_{q' \ell'}
		+
		\text{11 cyclic} .
\end{equation}

Now focus on the contracted terms
$\tilde{u}_{i' p' q'} \tilde{u}_{p' r' s'}$. Summing over the permutations
$\ell' \rightarrow \{ j', k' \}$ is equivalent to symmetrization over
$\{ q', r', s' \}$. Therefore this term can be combined with the
$\tilde{u}_{i' q' r' s'}$ source in~\eqref{eq:beta-source-integral},
giving a total contribution to the four-point function of the form
\begin{equation}
	\label{eq:g-NL-term}
	\beta_{\alpha' \beta' \gamma' \delta'}
	\supseteq
		\Gamma_{\alpha' q' r' s'}
		\Gamma_{\beta' j'}
		\Gamma_{\gamma' k'}
		\Gamma_{\delta' \ell'}
		\Sinit_{q' j'}
		\Sinit_{r' k'}
		\Sinit_{s' \ell'}
		+
		\text{3 cyclic} ,
\end{equation}
where we have defined $\Gamma_{\alpha' q' r' s'}$ to satisfy
\begin{equation}
	\label{eq:gamma-four-def}
	\begin{split}
	\Gamma_{\alpha' q' r' s'}
	= \mbox{}
	&
		\Gamma_{\alpha' i'}
		\int_{N_0}^N \d N' \;
		\tilde{u}_{i' q' r' s'}(N')
	\\ & \mbox{}
	+ \Big(
			\Gamma_{\alpha' i'}
			\int_{N_0}^N \d N' \;
			\tilde{u}_{i' p' q'}(N')
			\int_{N_0}^{N'} \d N'' \;
			\tilde{u}_{p' r' s'}(N'')
			+
			[ q' \rightarrow \{ r', s' \} ]
	\Big) .
	\end{split}
\end{equation}
As with the previous examples of
$\Gamma$-matrices,
the momentum dependence of
$\Gamma_{\alpha' q' r' s'}$ is a pure $\delta$-function.
It can be converted to a pure flavour
matrix by the rule
\begin{equation}
	\Gamma_{\alpha' q' r' s'} =
		(2\pi)^3
		\delta(\vect{k}_\alpha - \vect{k}_q - \vect{k}_r - \vect{k}_s)
		\Gamma_{\alpha q r s} .
\end{equation}
By explicit differentiation and back-substitution, it can be shown that
this flavour matrix satisfies the ordinary differential equation
\begin{equation}
	\label{eq:gamma-four-flow}
	\frac{\d\Gamma_{\alpha q r s}}{\d N}
	=
		u_{\alpha \beta} \Gamma_{\beta q r s}
		+
		\Big(
			u_{\alpha \beta \gamma} \Gamma_{\beta q r} \Gamma_{\gamma s}
			+
			\text{2 cyclic}
		\Big)
		+
		u_{\alpha \beta \gamma \delta}
			\Gamma_{\beta q}
			\Gamma_{\gamma r}
			\Gamma_{\delta s} .
\end{equation}
We have already seen that the lower-order Taylor coefficients
$\Gamma_{\alpha i}$ and $\Gamma_{\alpha ij}$
are determined by the evolution equations~\eqref{eq:jacobi-equation}
(with primed indices exchanged for unprimed ones)
and~\eqref{eq:gamma2-evolve};
for an extended discussion, see Ref.~\cite{Seery:2012vj}.
These equations provide an efficient means to compute the
``$\delta N$ coefficients'' numerically.

Returning to the four-point function,
we must also include the initial condition
\begin{equation}
	\label{eq:4pf-ic}
	\beta_{\alpha' \beta' \gamma' \delta'}
	\supseteq
		\Gamma_{\alpha' i'}
		\Gamma_{\beta' j'}
		\Gamma_{\gamma' k'}
		\Gamma_{\delta' \ell'}
		\Binit_{i' j' k' \ell'} .
\end{equation}
Repeating the steps described above, it can be shown that
\begin{equation}
	\Gamma_{\alpha q r s}
	=
		\frac{\partial \Gamma_{\alpha q r}}{\partial \varphi_s(N_0)}
	=
		\frac{\partial^3 \varphi_\alpha(N)}{\partial \varphi_q(N_0)
											\partial \varphi_r(N_0)
											\partial \varphi_s(N_0)} .
\end{equation}
Therefore
we have reproduced the usual Taylor expansion formulae for the
trispectrum.
Specifically, Eq.~\eqref{eq:4pf-ic} matches
(8) of Ref.~\cite{Seery:2006vu},
and
Eqs.~\eqref{eq:A-P-term}, \eqref{eq:tau-NL-term} and~\eqref{eq:g-NL-term}
match
(73), (74) and (75) of the same reference.
These expressions were later given in slightly more generality
by Byrnes, Sasaki \& Wands \cite{Byrnes:2006vq}.
In the formulation given by these authors,
Eqs.~\eqref{eq:A-P-term}, \eqref{eq:tau-NL-term}, \eqref{eq:g-NL-term}
and~\eqref{eq:4pf-ic}
of this paper
match (36) of Ref.~\cite{Byrnes:2006vq}.

\section{Transformation to the curvature perturbation}
\label{sec:gauge-transform}

We now have the transport equations which evolve the $n$-point
functions of the scalar field perturbations during inflation, up to
and including $n=4$.
These can be obtained either by solving the ``shape equations'',
Eqs.~\eqref{eq: g te}--\eqref{eq: tau te}, or using
Eq.~\eqref{eq:gamma-four-flow} to evolve the $\Gamma$-matrices.
For the latter case, the initial conditions are
$\Gamma_{\alpha i} = \delta_{\alpha i}$ at $N = N_0$,
with all other $\Gamma$-matrices zero there.

\subsection{Curvature perturbation at third order}

The scalar field fluctuations are not observable by themselves.
At present we have observational evidence only for a single
primordial fluctuation---the density fluctuation, which is a
nonlinear and model-dependent combination of the field fluctuations.
The appropriate combination can be deduced from the displacement
$\delta N$
(measured in e-folds) between
a fixed spatially-flat hypersurface and an adjacent
uniform-density hypersurface with which it coincides on average.
This displacement is determined by the field
configuration on the spatially
flat hypersurface.
Therefore
$\zeta = \delta N = \delta [N (\varphi_\alpha)]$, yielding%
\begin{equation}
	\label{eq:deltaN}
	\zeta
	=
		N_\alpha \delta \varphi_\alpha
		+
		\frac{1}{2!} N_{\alpha\beta}
		(
			\delta \varphi_\alpha \delta \varphi_\beta
			-
			\langle
				\delta \varphi_\alpha \delta \varphi_\beta
			\rangle
		)
		+
		\frac{1}{3!} N_{\alpha\beta\gamma}
		(
			\delta \varphi_\alpha \delta \varphi_\beta \delta \varphi_\gamma
			-
			\langle
				\delta \varphi_\alpha \delta \varphi_\beta
				\delta \varphi_\gamma
			\rangle
		)
		+
		\cdots
		,
\end{equation}
where
$N_\alpha = \partial N / \partial \varphi_\alpha$
and similarly for the higher derivatives.
Note that these are ordinary partial derivatives, with
all quantities evaluated at the same time: they are not the
nonlocal variational derivatives which appear in the
Lyth--Rodr\'{\i}guez Taylor expansion.
In particular, we are not using the $\delta N$ formula~\eqref{eq:deltaN}
to account for any time dependence of the correlation
functions; this is handled by the transport
equations.
Eq.~\eqref{eq:deltaN} is used solely to obtain the relationship
between the $\delta\varphi_\alpha$ and $\zeta$.
There are various other ways in which this could be obtained.
Malik \& Wands gave a comprehensive discussion~\cite{Malik:2008im}
from the viewpoint of traditional cosmological perturbation theory.
Another approach was used by Maldacena~\cite{Maldacena:2002vr}.
Eq.~\eqref{eq:deltaN} has the advantage that it computes the
transformation only in the superhorizon limit
$k/aH \rightarrow 0$, which is all we require.

Calculation of the derivatives
$N_\alpha$, $N_{\alpha\beta}$ and $N_{\alpha\beta\gamma}$
is tedious, although straightforward in principle.
Ref.~\cite{Seery:2012vj} used a raytracing method which gave the
relation a geometrical meaning.
It would be interesting to apply this technique at third order, but
it is helpful primarily for analytic and geometric
intuition rather than numerical optimization.
Ref.~\cite{Mulryne:2009kh} exploited the fact that
any potential is separable for first order displacements to set up
constants of the motion, as originally done by Garc\'{\i}a-Bellido
\& Wands~\cite{GarciaBellido:1995qq,Vernizzi:2006ve}. However, this method
is relatively lengthy even for the second-order coefficient
$N_{\alpha\beta}$.
Here we employ a simpler alternative.

We first focus on a \emph{single} trajectory
and measure the number
of e-folds $N$ accumulated along it.
During any period where the density decreases monotonically
we may measure $N$ as a function of $\rho$.
Consider the number of e-folds $\Delta N$ which elapse between
some arbitrary point on the trajectory
(the ``starting point'') and a nearby hypersurface of fixed
density $\rho_c$. Under the slow-roll approximation,
the density at the starting point is simply the potential energy
evaluated there. Therefore we may express $\Delta N$
as a Taylor expansion in the difference $\Delta \rho = \rho_c - V$,
\begin{equation}
	\label{eq:gauge-delta-N}
	\Delta N
	=
		N(V + \Delta \rho) - N(V)
		=
		\frac{\d N}{\d \rho} \Delta \rho
		+
		\frac{1}{2!} \frac{\d^2 N}{\d \rho^2}
		\Delta \rho^2
		+
		\frac{1}{3!} \frac{\d^3 N}{\d \rho^3}
		\Delta \rho^3
		+
		\cdots
		.
\end{equation}
Note that the differential coefficients are \emph{ordinary}
derivatives taken along the trajectory.
In
Eq.~\eqref{eq:gauge-delta-N} they are evaluated at the starting point.

We now perturb the starting point by an amount
$\delta \varphi_\alpha$
while keeping the final hypersurface fixed.
In general $\delta \varphi_\alpha$ will not be aligned with the
inflationary trajectory used to construct the $\rho$-derivatives in
Eq.~\eqref{eq:gauge-delta-N},
which therefore vary.
The same is true for the displacement $\Delta \rho$.
Accounting for both these effects changes the
total elapsed e-folds by an amount $\delta(\Delta N)$.
Finally, to study fluctuations around the hypersurface $\rho = \rho_c$
we take the limit $\Delta \rho \rightarrow 0$,
after which $\delta(\Delta N) \rightarrow \zeta$.
The advantage of this method is that it
uses the handful
of low-order derivatives appearing in~Eq.~\eqref{eq:gauge-delta-N}
to isolate the limited
information we require regarding local properties of the
transformation:
higher-order information is discarded at the outset.
This contrasts with the constants-of-motion approach used
in~Ref.~\cite{Mulryne:2009kh}, where
high-order information is implicitly kept through the majority of the
computation, although it is never used.

Under a shift of the starting point we conclude
\begin{equation}
	\delta ( \Delta \rho )
	=
		-
		V_\alpha \delta \varphi_\alpha
		-
		\frac{1}{2!}
		V_{\alpha\beta} \delta \varphi_\alpha \delta \varphi_\beta
		-
		\frac{1}{3!}
		V_{\alpha\beta\gamma} \delta \varphi_\alpha \delta \varphi_\beta
		\delta \varphi_\gamma
		-
		\cdots .
\end{equation}
By retaining contributions to $\rho$ from the kinetic
energy, and evaluating the
differential coefficients
in~\eqref{eq:gauge-delta-N} without use of the slow-roll approximation,
this approach could be extended to provide the transformation
from the full phase space variables $\delta\varphi_\alpha$,
$\delta \dot{\varphi}_\alpha$ to $\zeta$.
This was done in Ref.~\cite{Dias:2012nf}.

Invoking the slow-roll approximation, we may calculate
the derivative $\d N / \d \rho$,
\begin{equation}
	\left.
	\frac{\d N}{\d \rho}
	\right|
	=
		\frac{\d N}{\d t}
		\left. \frac{\d t}{\d V} \right|_{\gamma}
		=
		-
		\frac{3H^2}{V_\alpha V_\alpha}
		=
		-
		\frac{1}{\Mp^2} \frac{V}{V_\alpha V_\alpha},
\end{equation}
where $\d V/\d t$ is to be computed along the trajectory $\gamma$.
Higher derivatives can be obtained in the same way, by repeated
differentiation with respect to $t$ and use of the chain rule to
convert these into derivatives with respect to $\rho$. We obtain
\begin{align}
	\left. \frac{\d^2 N}{\d\rho^2} \right| & =
	- \frac{1}{\Mp^2} \left(
		\frac{1}{V_\lambda V_\lambda}
		-
		2 \frac{V V_\alpha V_\beta V_{\alpha\beta}}{(V_\lambda V_\lambda)^3}
	\right) ,
	\\
	\left. \frac{\d^3 N}{\d\rho^3} \right| & =
	\frac{1}{\Mp^2} \left(
		4 \frac{V_\alpha V_\beta V_{\alpha\beta}}{(V_\lambda V_\lambda)^3}
		-
		12 \frac{V ( V_\alpha V_\beta V_{\alpha\beta} )^2}
			{(V_\lambda V_\lambda)^5}
		+
		4 \frac{V V_\alpha V_{\alpha\beta} V_{\beta\gamma} V_\gamma}
			{(V_\lambda V_\lambda)^4}
		+
		2 \frac{V V_{\alpha\beta\gamma} V_\alpha V_\beta V_\gamma}
			{(V_\lambda V_\lambda)^4}
	\right) .
\end{align}

The first and second-order variations are
\begin{align}
	N_\alpha & = - \left. \frac{\d N}{\d \rho} \right|
		V_\alpha ,
	\\
	N_{\alpha\beta} & = - \left. \frac{\d N}{\d \rho} \right|
		V_{\alpha\beta}
	+ \left. \frac{\d^2 N}{\d \rho^2} \right| V_\alpha V_\beta
	+ \frac{1}{\Mp^2} \big( V_{\alpha} A_\beta + V_\beta A_\alpha \big)
	,
\end{align}
which agree with existing expressions in the literature.
(See below for the definition of $A_{\alpha}$.)
At third order we find
\begin{equation}
	\begin{split}
		N_{\alpha\beta\gamma}
		= \mbox{}
			&
			-
			\left. \frac{\d^3 N}{\d \rho^3} \right|
			V_\alpha V_\beta V_\gamma
			+
			\left(
				\left. \frac{\d^2 N}{\d \rho^2} \right|
				V_\alpha V_{\beta\gamma}
				+
				\text{cyclic}
			\right)
			-
			\left. \frac{\d N}{\d \rho} \right|
			V_{\alpha\beta\gamma}
		\\ & \mbox{}
			+ \frac{1}{\Mp^2}
			\left(
				A_{\alpha} V_{\beta\gamma}
				+
				\text{cyclic}
			\right)
			+
			\frac{1}{\Mp^2}
			\left(
				B_{\alpha\beta} V_\gamma
				+
				\text{cyclic}
			\right)
			+
			\frac{1}{\Mp^2}
			\left(
				C_{\alpha} V_\beta V_\gamma
				+
				\text{cyclic}
			\right) .
	\end{split}
\end{equation}
The tensors $A_\alpha$, $B_{\alpha\beta}$
and $C_\alpha$ have been defined to satisfy
\begin{align}
	A_\alpha
	& =
		\frac{V_\alpha}{V_\lambda V_\lambda}
		-
		2
		\frac{V V_\kappa V_{\kappa \alpha}}{(V_\lambda V_\lambda)^2}
	\\
	B_{\alpha\beta}
	& =
		\frac{V_{\alpha\beta}}{V_\lambda V_\lambda}
		-
		2
		\frac{V_\kappa V_{\kappa\alpha} V_\beta
			+ V_\kappa V_{\kappa\beta} V_\alpha}{(V_\lambda V_\lambda)^2}
		+
		8
		\frac{V V_\kappa V_{\kappa \alpha} V_{\epsilon} V_{\epsilon \beta}}
			{(V_\lambda V_\lambda)^3}
		-
		2
		\frac{V V_{\kappa \alpha} V_{\kappa \beta}}{(V_\lambda V_\lambda)^2}
		-
		2
		\frac{V V_\kappa V_{\kappa \alpha \beta}}{(V_\lambda V_\lambda)^2}
	\\
	C_\alpha
	& =
		\frac{V_{\alpha\beta}}{V_\lambda V_\lambda}
		+
		2
		\frac{V_\alpha V_\kappa V_{\kappa\epsilon} V_\epsilon}
			{(V_\lambda V_\lambda)^3}
		-
		12
		\frac{V V_\kappa V_{\kappa \epsilon} V_\epsilon V_\rho V_{\rho\alpha}}
			{(V_\lambda V_\lambda)^3}
		+
		4
		\frac{V V_\kappa V_{\kappa\epsilon} V_{\epsilon\alpha}}
			{(V_\lambda V_\lambda)^3}
		+
		2
		\frac{V V_\kappa V_{\kappa\epsilon\alpha} V_\epsilon}
			{(V_\lambda V_\lambda)^3}
\end{align}

\subsection{Inflationary observables}
\label{sec:tau-g-formulae}

Finally, we must assemble all these contributions to obtain expressions
for $\tauNL$ and $\gNL$. We find
\begin{align}
	P_\zeta
	& =
		N_\alpha N_\beta \Sigma_{\alpha\beta}
	\label{eq:Pzeta}
	\\
	\frac{6}{5} \fNL
	& =
		\frac{N_\alpha N_\beta N_\gamma \alpha_{\alpha\mid\beta\gamma}
		+ N_{\alpha\beta} N_\gamma N_\delta \Sigma_{\alpha\gamma}
		\Sigma_{\beta\delta}}
		{(N_\omega N_\zeta \Sigma_{\omega\zeta})^2}
	\label{eq:fNL}
	\\
	\tauNL
	& =
		\frac{N_\alpha N_\beta N_\gamma N_\delta
		\tau_{\alpha\beta\mid\gamma\delta}
		+
		2
		N_\alpha N_\beta N_\gamma N_{\lambda\mu}
		a_{\alpha\mid\beta\lambda} \Sigma_{\gamma\mu}
		+
		N_{\alpha\beta} N_{\gamma\delta} N_{\lambda} N_{\mu}
		\Sigma_{\alpha\gamma} \Sigma_{\beta\lambda} \Sigma_{\delta\mu}}
		{(N_\omega N_\zeta \Sigma_{\omega\zeta})^3}
	\label{eq:tauNL}
	\\
	\frac{54}{25} \gNL
	& =
		\frac{N_\alpha N_\beta N_\gamma N_\delta
		g_{\alpha\mid\beta\gamma\delta}
		+
		3
		N_\alpha N_\beta N_\gamma N_{\lambda\mu}
		a_{\lambda\mid\alpha\beta} \Sigma_{\gamma\mu}
		+
		N_{\alpha\beta\gamma} N_\delta N_\lambda N_\mu
		\Sigma_{\alpha\delta} \Sigma_{\beta\lambda} \Sigma_{\gamma\mu}}
		{(N_\omega N_\zeta \Sigma_{\omega\zeta})^3} .
	\label{eq:gNL}
\end{align}

\section{Alternative approaches}
\label{sec:backwards}

In this paper, our approach to calculating the statistics of the curvature perturbation has been to develop 
transport equations for objects such as the $n$-point functions
[Eqs.~\eqref{eq: 2pf te}, \eqref{eq: 3pf te} and~\eqref{eq: 4pf te}], 
or their shape tensors
[Eqs.~\eqref{eq: 2pf flavour},~\eqref{eq: 3pf flavour} and~\eqref{eq: g te}--\eqref{eq: tau te}].
The results of~\S\ref{sec:taylor} show that this is equivalent
to the Lyth--Rodr\'{\i}guez Taylor expansion
\begin{equation}
	\delta \varphi_\alpha =
		\Gamma_{\alpha i} \delta \varphi_i
		+
		\frac{1}{2!} \Gamma_{\alpha i j} \delta \varphi_i \delta \varphi_j
		+
		\frac{1}{3!} \Gamma_{\alpha i j k} \delta \varphi_i \delta \varphi_j
			\delta \varphi_k
		+ \cdots ,
	\label{eq:Gexpn}
\end{equation}
where we recall that objects with Greek indices are evaluated at time $N$,
and those with Latin indices at some earlier time $N_0$, which is usually
taken as the common time of horizon exit for the $\vect{k}$-modes under
consideration.%
	\footnote{Indeed, one can verify that inserting~\eref{eq:Gexpn}
	into~\eref{eq:jacobi}
	and equating coefficients order-by-order reproduces
	the $\Gamma$-matrix evolution equations with unprimed indices
	[Eqs.~\eref{eq:jacobi-equation}, \eref{eq:gamma2-evolve} 
	and~\eref{eq:gamma-four-flow}].}
	
\para{`Forward' and `backward' methods}%
To solve these equations
we must supply a boundary condition at $N = N_0$, but we are free
to choose how this is done:
we may start either with $N_0$ at the initial epoch and evolve $N$
\emph{forward} to the time of interest,
or fix $N$ at this time and evolve $N_0$ \emph{backwards}.
These approaches are distinct but equally valid, because
(at least during inflation)
there is no obstacle to computing the relevant initial conditions
at any time of our choosing.
The transport equations we have described in this paper are of the
forwards variety.

Eq.~\eqref{eq:Gexpn} shows explicitly what must be computed in order
to completely characterize the fluctuations $\delta \varphi_\alpha$
at any given order. At first order in a $d$-field
slow-roll model, we require the $d^2$ independent
components of the Jacobi map $\Gamma_{\alpha i}$.
At second order there are $d^3$ components of $\Gamma_{\alpha ij}$,
reduced to $d^2(d+1)/2$ after accounting for symmetries.
Finally, at third order there are $d^4$ components
of $\Gamma_{\alpha ijk}$, which reduce to $d^2(d+1)(d+2)/6$
independent components
after symmetries.
We conclude that to compute all two-point functions
in such a model
requires solution of $\Or(d^2)$ differential equations.
Likewise,
all three-point functions requires $\Or(d^3)$ equations,
and all four-point functions requires $\Or(d^4)$ equations.
This is to be expected, because there are $\Or(d^m)$ independent
$m$-point functions.

\para{Autocorrelation functions of $\zeta$ only}%
Sometimes we do not require all correlation functions, but only the
autocorrelation functions of $\zeta$.
In such cases it would be 
advantageous if an autonomous set
of transport equations could be set up for the
Taylor coefficients of $\zeta$ rather than $\delta \varphi_\alpha$,
\begin{equation}
	\delta N
		=
			N_{i} \delta \varphi_i
			+
			\frac{1}{2!} N_{i j} \delta \varphi_i \delta \varphi_j
			+
			\frac{1}{3!} N_{i j k} \delta \varphi_i \delta \varphi_j
										  \delta \varphi_k .
\end{equation}
This would require the solution of only $\Or(d^{m-1})$ independent
equations to obtain the $m$-point function of $\zeta$.
This saving could be helpful in models with a large number of fields.
There is currently no \emph{forwards} formulation of this type
but
a set of \emph{backwards} equations were given by
Yokoyama, Suyama \& Tanaka~\cite{Yokoyama:2007uu, Yokoyama:2007dw},
and later extended to the 
trispectrum~\cite{Yokoyama:2008by}.

The Taylor coefficients for $N$ can be expressed in terms of the
$\Gamma$-matrices,
\begin{align}
	N_i 
	& =
		N_\alpha \Gamma_{\alpha i}
	\label{eq:n-one}
	\\
	N_{ij}	
	& =
		N_\alpha \Gamma_{\alpha i j}+ N_{\alpha \beta} \Gamma_{\alpha i} \Gamma_{\beta j}
	\label{eq:n-two}
	\\	
	N_{ijk} 
	& =	
		N_\alpha \Gamma_{\alpha i j k} + N_{\alpha \beta \gamma} \Gamma_{\alpha i} \Gamma_{\beta j} \Gamma_{\gamma k}
		+
		\left(
			N_{\alpha \beta}  \Gamma_{\beta i} \Gamma_{\alpha j k} +
			\text{2 cyclic}
		\right) .
	\label{eq:n-three}
\end{align}
We could attempt to obtain forward
transport equations by direct differentiation with
respect to time followed by use of the $\Gamma$-matrix evolution equations.
But this does not generate a closed set of autonomous equations
because derivatives of the $N_{\alpha \cdots}$ also appear,
which obstruct an attempt to eliminate the $\Gamma$-matrices in favour
of their $N$ counterparts.

Instead,
the backwards equations of Yokoyama~et~al. can be derived as follows.
As described above, we fix $N$ to be the late time of interest
and aim to evolve $N_0$ backwards.
The backwards evolution of $\Gamma_{\alpha i}$ can be obtained
very simply
by differentiating~\eqref{eq:Gexpn} while
keeping $\delta \varphi_\alpha$ fixed,
or alternatively 
by differentiating~\eqref{eq:gamma-one} with respect to $N_0$.
Whichever method is chosen, we find
$\d \Gamma_{\alpha i} / \d N_0 = - \Gamma_{\alpha j} u_{ji}$.
Subsequently differentiating~\eqref{eq:n-one} with respect to $N_0$
and using this relation, we obtain an autonomous set of equations for
$N_i$,
\begin{equation}
	\frac{\d N_i}{\d N_0} = - N_j u_{ji} .
	\label{eq:NI}
\end{equation}
This technique can be extended to higher orders, giving evolution
equations for $N_{ij}$ and $N_{ijk}$.%
	\footnote{The evolution equations for $\Gamma_{\alpha ij}$
	and $\Gamma_{\alpha ijk}$, which are required to obtain these
	results, are
	\begin{align}
		\frac{\d \Gamma_{\alpha ij}}{\d N_0}
			& = 
			- \Gamma_{\alpha m} u_{mij}
			- \Gamma_{\alpha im} u_{mj}
			- \Gamma_{\alpha mj} u_{mi} \\
		\frac{\d \Gamma_{\alpha ijk}}{\d N_0}
			& =
			- \Gamma_{\alpha m} u_{mijk}
			- \Big(
				\Gamma_{\alpha im} u_{mjk}
				+
				\Gamma_{\alpha mjk} u_{mi}
				+
				\text{2 cyclic}
			\Big) .
	\end{align}}
We find
\begin{equation}
	\frac{\d N_{ij}}{\d N_0} = - N_k u_{kij} - N_{ik} u_{kj} - N_{jk} u_{ki} ,
\end{equation}
and 
\begin{equation}
	\frac{\d N_{ijk}}{\d N_0}
	=
		- N_{\ell} u_{\ell ijk}
		- \left(
			N_{i \ell} u_{\ell jk}
			+
			N_{jk \ell} u_{\ell i}
			+
			\text{2 cyclic}
		\right) .
\end{equation}	
The first of these was given in Ref.~\cite{Seery:2012vj}.
Here we have extended the method to include $N_{ijk}$, which enables
trispectrum quantities to be calculated.
These equations should be solved
with initial
conditions chosen so that $N_i$, $N_{ij}$ and $N_{ijk}$
equal the transformation matrices $N_{\alpha}$, $N_{\alpha\beta}$
and $N_{\alpha\beta\gamma}$, respectively, at $N_0 = N$.

If we require only the bispectrum of $\zeta$
and
are prepared to take the field fluctuations at time $N_0$
to be Gaussian and uncorrelated with each other, then more is possible.
Under these circumstances, Yokoyama~et~al. showed that the $\Or(d^2)$
equations for $N_{ij}$ could be replaced by only $\Or(d)$ equations
for an auxiliary quantity $\Theta_\alpha = \Gamma_{\alpha i} N_i$
\cite{Yokoyama:2007dw,Yokoyama:2007uu}.

\para{Constraint for first-order coefficients}%
There is a further simplification which can be made
for the $N_i$ system.
Using the flow equation
$\d \phi_i = u_i \, \d N$,
it follows that the displacement $\d \phi_i = u_i$
precisely tangent to the trajectory
generates a
change in the e-foldings required to reach the final
uniform density slice corresponding to
\begin{equation}
	\delta N = u_i N_i = -1 .
	\label{eq:dN-constraint}
\end{equation}
This implies that \emph{one} of the $N_i$ can be determined algebraically
in terms of the others, without solving a separate differential equation.
Therefore, in a two-field model, the Yokoyama~et~al. equations~\eqref{eq:NI}
can be decoupled,
\begin{equation}
	\frac{\d N_\phi}{\d N_0} = \left(
		\frac{u_\phi}{u_\chi} u_{\chi \phi} - u_{\phi\phi}
	\right) N_\phi
	+
	\frac{u_{\chi\phi}}{u_\chi} ,
	\label{eq:mw-equation}
\end{equation}
where we have labelled the fields $\phi$ and $\chi$.
A similar equation can be given for $N_\chi$, but it is unnecessary
because~\eqref{eq:dN-constraint} can be used to obtain $N_\chi$
once $N_\phi$ is known.
Although
the possibility of decoupling these equations is interesting,
it confers no particular advantages.

A variation of the Yokoyama~et~al. formulation was recently given by
Mazumdar \& Wang~\cite{Mazumdar:2012jj}
in which they pointed out the possibility of this decoupling
in the two-field case,
although without making explicit use of the
constraint~\eqref{eq:dN-constraint}.
Their analysis is
equivalent to the one presented here,
and in Appendix A of Ref.~\cite{Seery:2012vj}.
Mazumdar \& Wang ascribed the possibility of decoupling to the choice of
coordinates used in their derivation.
However, the evolution equation~\eqref{eq:NI}
can be derived using any convenient method and is independent of such choices.
The argument above shows that decoupling is
a consequence of the constraint~\eqref{eq:dN-constraint}, and is
a special feature of the two-field system.
In a general $d$-field model, the best that can be obtained is
a coupled system of $d-1$ equations.

\section{Numerical results}
\label{sec:results}

We now illustrate the transport approach using a number of concrete models. 
For each model, we numerically solve
Eqs.~\eref{eq: 2pf flavour},~\eqref{eq: 3pf flavour}
and~\eqref{eq: g te}--\eqref{eq: tau te}, 
and use Eqs.~\eqref{eq:fNL}--\eqref{eq:gNL}
to determine the values
of $\fNL$, $\tauNL$ and $\gNL$ from horizon crossing
onwards.
We label the number of e-folds of inflation from $N=0$ at horizon exit.

\subsection{Numerical Examples}

\para{D-brane model}%
Our first example was studied by Dias, Frazer \& Liddle~\cite{Dias:2012nf}.
It is an approximation to
inflation driven by the motion of a D-brane in a warped throat,
allowing for angular degrees of freedom. 
In that study, the authors employed the transport approach to calculate
the distribution of
observable parameters over a large number of realizations of their model.
However,
they 
restricted attention to the spectrum and local-type bispectrum.   
Here we present the evolution of the 
local-type trispectrum parameters for one typical realization.
 
\begin{figure}[htb]
\centering
\includegraphics[width=0.8\textwidth]{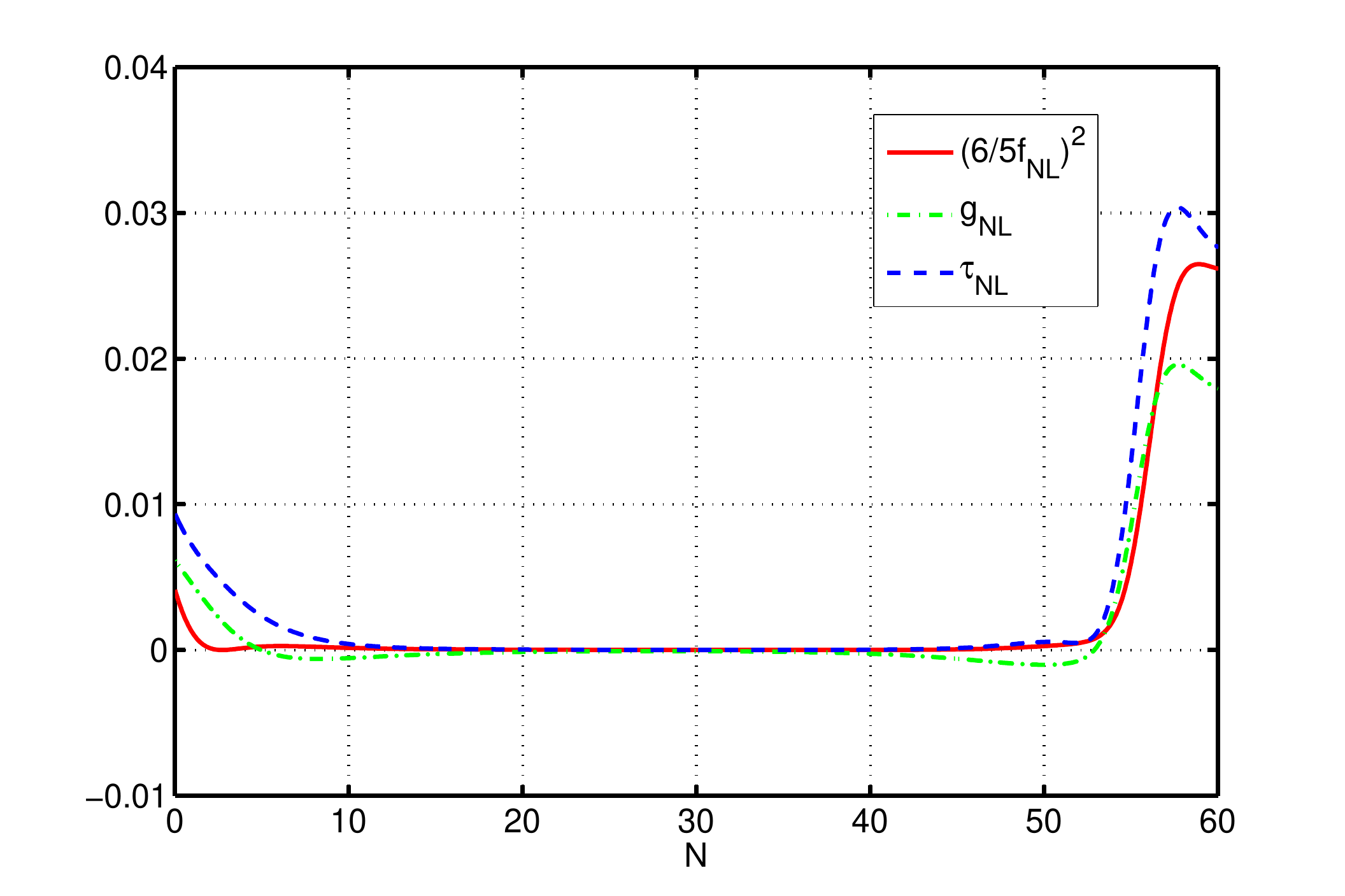}
\caption{Evolution of $\tauNL$, $\gNL$ and $(6/5 \fNL)^2$ for the
inflexion-point potential~\eqref{eq:inflexion-point}.
Initial conditions and parameter choices are described in the
main text.}
\label{fig:dbrane}
\end{figure}

The potential 
is given by 
\begin{equation}
	V = \alpha_0 + \alpha_1 \phi_1 + \alpha_3 \phi_1^3 + \beta \phi_2 ,
	\label{eq:inflexion-point}
\end{equation}
which contains an inflexion point
in the $\phi_1$ direction. Inflation occurs close to this inflexion point. 
We choose $\alpha_0 = 100 M^2 \Mp^2$, $\alpha_1 = M^2 \Mp$, 
$\alpha_3 = 5 M^2/\Mp$, $\beta = 5 M^2 \Mp$, $\phi_{1\text{exit}} = 0.5 \Mp$, and $\phi_{2\text{exit}} = 0.5 \Mp$, 
where the subscript `exit' indicates these are the initial values of the fields at horizon exit.  $M$ is an overall 
normalisation, which can be fixed to match the WMAP normalization of
the power spectrum.
These initial conditions have been
chosen to give $60$-efolds of inflation,
taking inflation to end when $\epsilon=1$.
Allowing the system to evolve past this point would lead to 
erroneous results
because we are employing slow-roll equations of motion.
As explained in~\S\ref{sec:transport} this could be
resolved by writing transport equations in 
the full phase-space. However, for simplicity, we do not 
do so here.
In Fig.~\ref{fig:dbrane} we give 
the evolution of $\tauNL$, $\gNL$, and $(6\fNL/5)^2$ for this choice of parameters and initial conditions.

For single-field models we recall that $\tauNL =(6\fNL/5)^2$,
which is relaxed to an inequality in multiple-field models
\cite{Suyama:2007bg,Smith:2011if}.
The use of the relative magnitude of $\tauNL$ and $(6\fNL/5)^2$ as
a diagnostic of the spectrum of active fields during inflation was
emphasized by Smidt~et~al.~\cite{Smidt:2010ra}, who made a forecast of
observational prospects.
Very recently, Assassi~et~al.~\cite{Assassi:2012zq}
gave precise formulae in terms of the spectrum of single-particle states.
This signature of multiple active fields is clearly visible
in Fig.~\ref{fig:dbrane}, although in this realization
the nongaussian parameters are too small to be observable.
(As a point of principle an inflexion point 
potential may give rise to a large local
bispectrum~\cite{Elliston:2011et} and
trispectrum~\cite{Elliston:2012wm}, via
the hilltop mechanism suggested by Kim~et~al.~\cite{Kim:2010ud}.
However, an observable signal can usually be obtained only 
for finely tuned initial conditions and parameter choices.)

\para{Quadratic-exponential model}%
Our second example was constructed by 
Byrnes et~al.~\cite{Byrnes:2008wi} as an example of a product-separable 
model which could give rise to a large $\fNL$ for
finely-tuned initial conditions. It was later studied by 
Elliston~et~al.~\cite{Elliston:2011dr} and
Huston~et~al.~\cite{Huston:2011fr}. 

\begin{figure}[htb]
\centering
\includegraphics[width=0.8\textwidth]{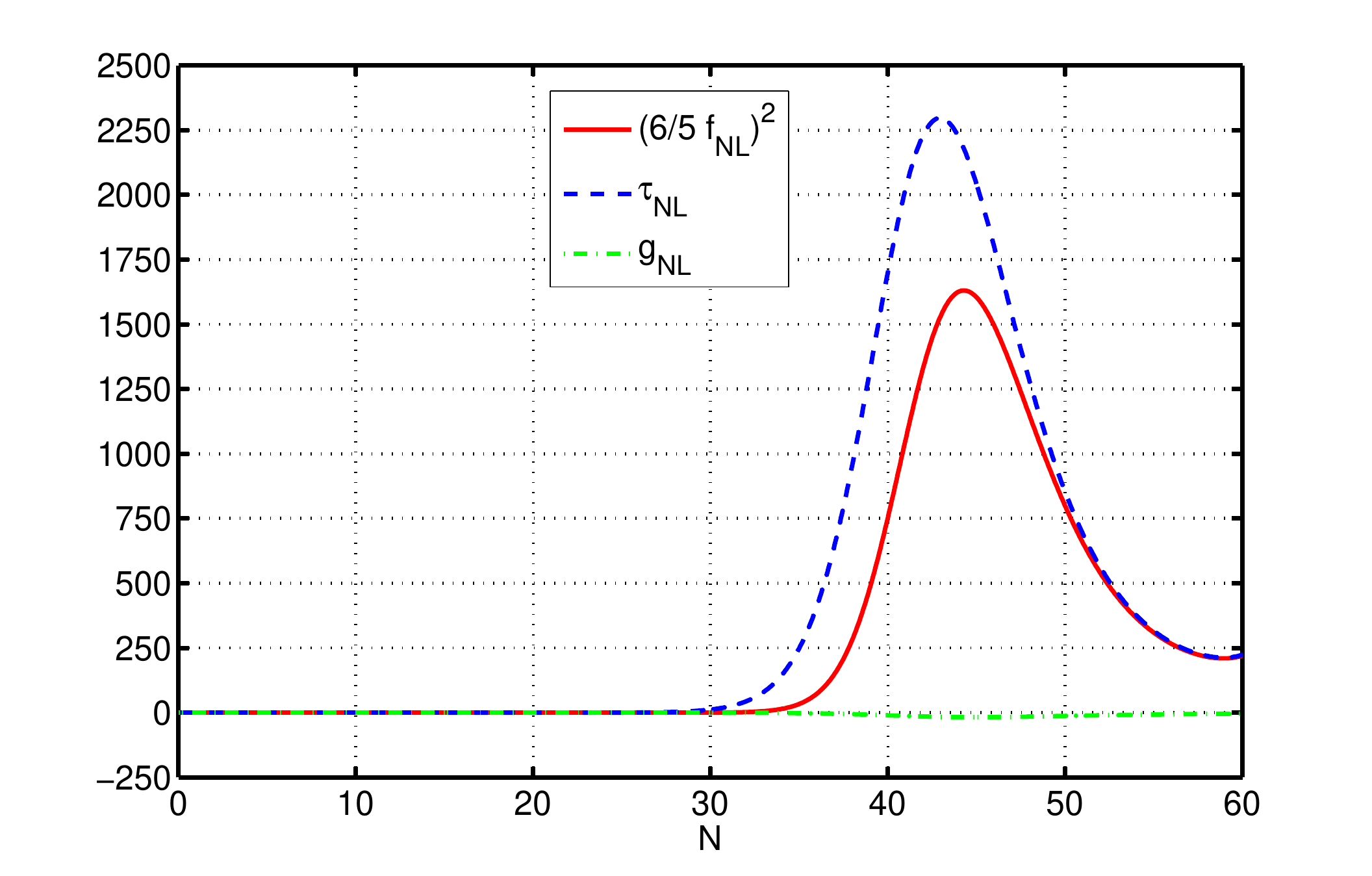}
\caption{Evolution of $\tauNL$, $\gNL$ and $(6/5 \fNL)^2$ 
for the potential~\eqref{eq:quadratic-exp}.
Initial conditions and parameter choices are described in the
main text.}
\label{fig:Chris}
\end{figure}

The potential is
\begin{equation}
	V= M^4 \phi_1^2 e^{-\lambda \phi_2^2} .
	\label{eq:quadratic-exp}
\end{equation}
We choose the parameter values and initial conditions
$\lambda = 0.05/\Mp^2$, $\phi_{1\text{exit}} = 16 \Mp$, 
and $\phi_{2\text{exit}}=0.001 \Mp$,
and fix $M$ as before to match the WMAP normalization.
These 
initial values also give $60$-efolds of inflation.
They have been chosen to select
a background trajectory which
gives rise to significant
nongaussianity.
In Fig. \ref{fig:Chris}, we present the evolution 
of the $\tauNL$ and $\gNL$ parameters in this model for the first time.
We also show the evolution of $(6\fNL/5)^2$.
However, although
the $\fNL$ and $\tauNL$ parameters are large at the end of inflation,
it is important to note that
the fluctuations are still evolving at this time.
Therefore the model is not predictive by itself:
it must be supplemented by post-inflationary evolution, which
tracks the fluctuations until the surface of last scattering,
or explains how all isocurvature modes eventually decay.

\para{Non-separable hybrid model}%
Finally, we present results for a hybrid-type potential
in the large field 
regime studied by Mulryne, Orani \& Rajantie~\cite{Mulryne:2011ni}.
This is an example of a non-separable potential.
For general initial conditions,
no analytic estimate is known for any of $\fNL$, $\tauNL$ or $\gNL$,
even assuming slow-roll.
Therefore
numerical 
methods, such as our implementation of the transport equations, become essential. The potential contains a 
hilltop region, and parameter choices and initial conditions can be chosen
so that the model is 
of the type discussed by Kim~et~al.~\cite{Kim:2010ud}.
This gives rise to large nongaussianity 
for initial conditions sufficiently close to the hilltop. 

\begin{figure}[htb]
\centering
\includegraphics[width=0.8\textwidth]{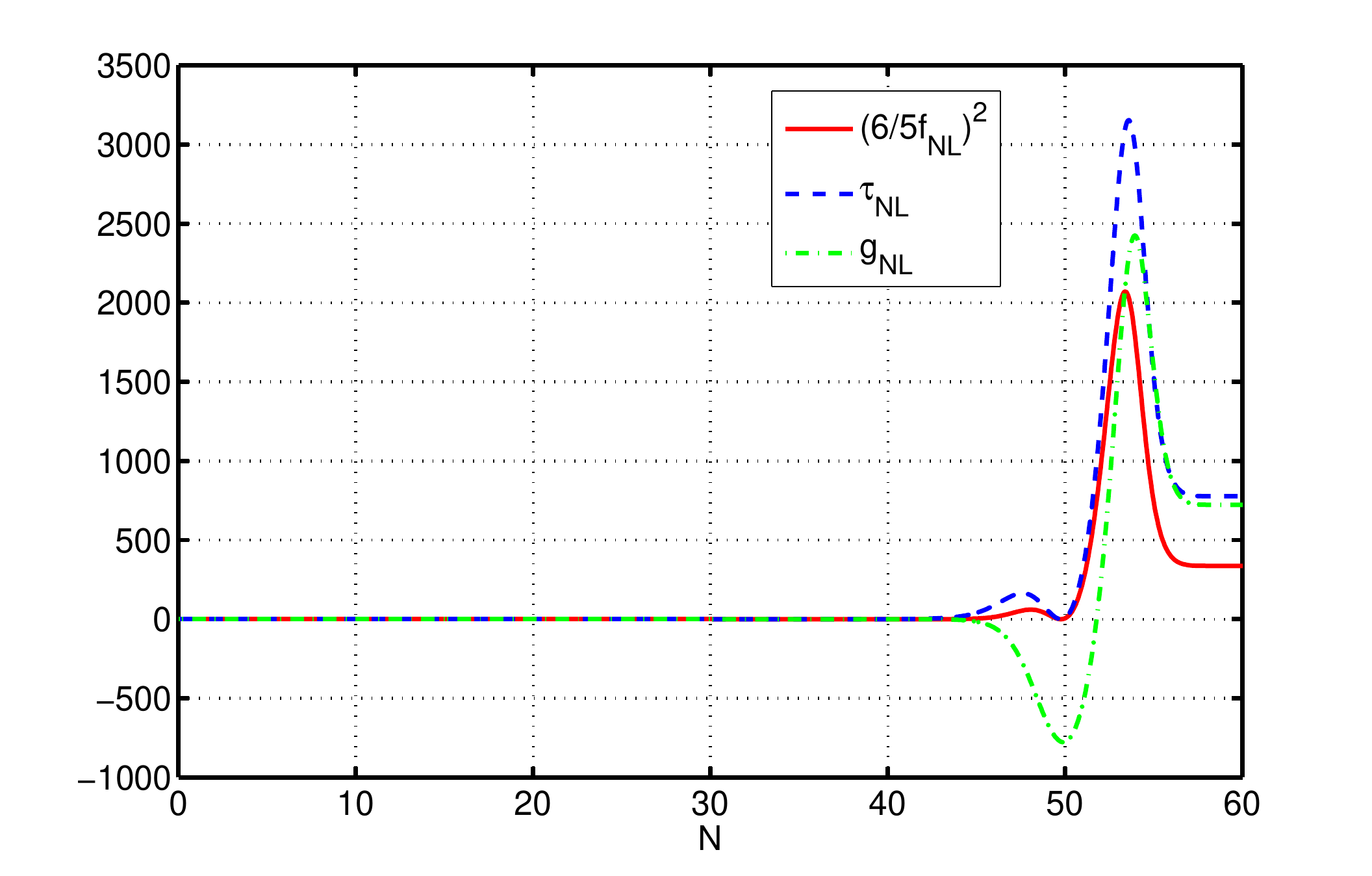}
\caption{Evolution of $\tauNL$, $\gNL$ and $(6/5 \fNL)^2$ 
for the potential~\eqref{eq:nonsep-model}.
Initial conditions and parameter choices are described in the
main text.}
\label{fig:Stefano}
\end{figure}

The potential satisfies
\begin{equation}
	V= M^4
		\left [
			\frac{1}{2} m^2 \phi_1^2
			+
			\frac{1}{2}g^2\phi_1^2 \phi_2^2
			+
			\frac{\lambda }{4} \left ( \phi_2^2-v^2 \right)^2
		\right ] ,
	\label{eq:nonsep-model}
\end{equation}
and
we choose the parameter values 
$g^2=v^2/\phi_{\text{crit}}^2$, $m^2=v^2$, $v=0.2\Mp$, $\phi_{\text{crit}}=20 \Mp$ and $\lambda=5$.
The initial conditions are $\phi_{1\text{exit}} = 15.5 \Mp$
and $\phi_{2\text{exit}} = 0.0015 \Mp$.
As above, these 
initial values give $60$-efolds of inflation
and have been adjusted to produce significant nongaussianity.
$M$ is adjusted as before.
In Fig.~\ref{fig:Stefano}, we present the evolution 
of the $\tauNL$ and $\gNL$ parameters in this model for the first time. In contrast to the previous example, 
the statistics here approach constant 
values before the end of inflation, reflecting the fact that isocurvature modes decay. We also give the evolution of $(6\fNL/5)^2$. 
\section{Discussion and Conclusions}
\label{sec:conclusions}

In this paper we have
provided transport equations to evolve the four-point functions
of a collection of light scalar fields during an inflationary phase.
The transport system can be thought of as a form of Boltzmann hierarchy,
and can be solved by similar methods.
Since inflationary fluctuations are typically close to Gaussian,
connected
correlation functions of increasing order are typically decreasing
in amplitude.
Therefore only a few low-order functions
are important in sourcing
those of higher order.
Truncating the hierarchy to include only these sources
generates the local-type ``squeezed'' and ``collapsed''
configurations.
We parametrize the amplitude of these configurations with
``shape tensors'' for which we have supplied evolution equations.
Expressing the correlation
functions of $\zeta$ in terms of those of the $\delta \varphi_\alpha$,
it is possible to extract $\tauNL$ and $\gNL$.
This analysis was given in \S\ref{sec:transport}.

This method of integrating the transport hierarchy expresses the
correlation functions in terms of the Jacobi fields
generated by the underlying phase space flow, and their derivatives.
One can regard this as a statement of the separate universe approximation.
The ``Jacobi map'' relates these fields to the variation of
a general solution of the equations of motion
with respect to its constants of integration.
Using this equivalence, we have shown that the result reproduces
the familiar Taylor expansion used by Lyth \& Rodr\'{\i}guez.
The procedure can be viewed as an application of classical
Hamilton--Jacobi theory.

Our equations supply a toolkit which can be used to study the
evolution of inflationary observables in any multi-field model
of interest, provided all fields possess canonical kinetic terms.
There are two equivalent approaches.
First, one can solve Eqs.~\eqref{eq: 2pf flavour},
\eqref{eq: 3pf flavour}, \eqref{eq: g te} and~\eqref{eq: tau te}
for the shape tensors corresponding to the two-, three- and
four-point functions, using suitable initial conditions.
Eqs.~\eqref{eq:Pzeta}--\eqref{eq:tauNL} can then be used to
extract observables.
Alternatively, one can solve the evolution equations
\eqref{eq:jacobi-equation} (after exchanging primed for unprimed
indices), \eqref{eq:gamma2-evolve}
and~\eqref{eq:gamma-four-flow}
for the Taylor coefficients of the ``$\delta N$ formalism'',
applied to the field fluctuations.
Once these are known, Eqs.~\eqref{eq:n-one}--\eqref{eq:n-three}
can be used to exchange them for the Taylor coefficients of $N$
itself. The usual formulae then allow observables
to be computed.
If the spectral index or its running are required,
they can be extracted using the methods described by
Dias~et~al.~\cite{Dias:2011xy,Dias:2012nf}.

Assuming slow-roll,
either method requires the solution of $\Or(d^m)$ equations
to obtain the $m$-point functions of a $d$-field model.
Since there are $\Or(d^m)$ independent correlation functions
it will not be possible to reduce this asymptotic complexity.
But
if only the autocorrelation functions of $\zeta$ are required,
then it may be advantageous to use the `backwards' formalism
introduced by Yokoyama, Suyama \& Tanaka,
in which one can reduce the number of equations to be solved
to $\Or(d^{m-1})$ by forfeiting the possibility
of obtaining correlation functions with insertions of isocurvature
modes. [For clarity,
we emphasize that
the formalism of Yokoyama~et~al. correctly
accounts for the influence of these isocurvature modes on the
evolution of the $\zeta$ correlation functions.
But it is not possible to determine mixed correlation functions, such
as $\langle \zeta s \rangle$, where $s$ is a field space
direction orthogonal to $\zeta$.]
Unfortunately,
it is often necessary to know something about such correlation
functions
to determine whether unquenched isocurvature modes
remain, which could change the inflationary prediction by
transferring their energy to the curvature fluctuation during or
after reheating. (We refer to Ref.~\cite{Seery:2012vj}
for a more comprehensive discussion.)
But in some cases
this may not be a concern,
and where this is true
our
extension of the formalism of Yokoyama~et~al.
allows trispectrum parameters to be obtained.

\ack
GJA was supported by the Science and Technology Facilities Council
[grant number ST/].
DJM acknowledges support from the Science and Technology Facilities Council
[grant number ST/J001546/1].
DS acknowledges support from the Science and Technology Facilities Council
[grant number ST/I000976/1]
and the Leverhulme Trust.
We would like to thank Lingfei Wang
and Antony Lewis for helpful discussions.

\appendix

\section{Contributions to the four-point function from the initial condition
of the three-point function}
\label{appendix:ic}

In this appendix we verify the claim made in \S\ref{sec:local-separation},
that the arbitrary initial condition for the three-point function
makes a negligible contribution to the sourced component of the
four-point function. In the text, this
was used to conclude that the initial value need not be retained in
Eqs.~\eqref{eq: g te}--\eqref{eq: tau te}.

It was first proved by Lyth \& Zaballa
that the initial condition for the three-point function could be neglected
in comparison with the sourced contribution whenever the sum of the two
was large enough to be observed \cite{Lyth:2005qj}.
Their argument was later simplified by Vernizzi \& Wands
\cite{Vernizzi:2006ve}.
The same result for the four-point function
follows from the analysis of
Refs.~\cite{Seery:2006vu,Seery:2008ax}.
However, we are unaware of a similar demonstration for the
question addressed in this appendix---the contribution of initial
value of the \emph{three}-point function
to the sourced component of the \emph{four}-point function.

We work with the variational formulation of the separate universe
approximation, as discussed by Lyth \& Rodr\'{\i}guez \cite{Lyth:2005fi}.
We write
\begin{equation}
	\zeta
	=
		N_{i} \delta \varphi_i
		+
		\frac{1}{2!} N_{ij}
		\delta \varphi_i \delta \varphi_j
		+
		\cdots ,
\end{equation}
where the Latin indices $i$, $j$, \ldots, have the same meaning
as in the main text.
We define the trispectrum $T_\zeta$ to be the four-point function with
its momentum-conservation $\delta$-function stripped away,
\begin{equation}
	\langle
		\zeta(\vect{k}_1)
		\zeta(\vect{k}_2)
		\zeta(\vect{k}_3)
		\zeta(\vect{k}_4)
	\rangle
	=
		(2\pi)^3 \delta(\vect{k}_1 + \vect{k}_2 + \vect{k}_3 + \vect{k}_4)
		T_\zeta .
\end{equation}
Using the initial value of the
three-point function computed in Ref.~\cite{Seery:2005gb},
the corresponding contribution to the sourced part of $T_\zeta$
can be written
\begin{equation}
	\begin{split}
		T_\zeta \supseteq N_i N_j N_k N_{mn}
		\Bigg\{
			&
			\delta^{im}
			\frac{H_\ast^2}{2k_1^3} \frac{H_\ast^4}{8 k_2^3 k_3^3 k_{14}^3}
			\sum_{\text{perms}}
			\frac{\dot{\varphi}_j \delta_{kn}}{2H_\ast}
			\Big(
				{- 3} \frac{k_3^2 k_{14}^2}{k_t}
				- \frac{k_3^2 k_{14}^2}{k_t^2} (k_2 + 2 k_{14})
				+ \frac{k_2^3}{2}
				- k_2 k_3^2
			\Big)
			\\ & \mbox{}
			+ [
				\vect{k}_1 \rightarrow \{ \vect{k}_2, \vect{k}_3 \}
			]
		\Bigg\}
		\\ & \mbox{}
			+ \text{cyclic permutations $\vect{k}_4 \rightarrow
				\{ \vect{k}_1, \vect{k}_2, \vect{k}_3 \}$}
	\end{split}
\end{equation}
where ``$\ast$'' denotes evaluation at horizon exit,
$k_t = k_1 + k_2 + k_{14}$,
the summation is over all simultaneous permutations
of the index set $\{ \beta, \gamma, \epsilon \}$
and the momenta $\{ \vect{k}_2, \vect{k}_3, \vect{k}_{14} \}$,
and
we have defined $k_{14} = |\vect{k}_1 + \vect{k}_4|$.

This contribution can be divided into an effective $\gNL$,
an effective $\tauNL$, and an `equilateral-type' term which does not
fit naturally into either of the local-type shapes.
The effective $\gNL$ can be written
\begin{equation}
	\label{eq:effective-g}
	\Delta \gNL = \frac{25}{72}
	\frac{N_i N_{ij} \dot{\varphi}_j / H_\ast}
	{(N_k N_k)^2}
\end{equation}
(the placement of indices is immaterial in this and other
expressions, since contraction occurs under the Kronecker-$\delta$),
and the effective $\tauNL$ is
\begin{equation}
	\label{eq:effective-tau}
	\Delta \tauNL = - \frac{1}{2} \frac{N_i N_{ij} N_j}
		{(N_k N_k)^3} .
\end{equation}
These expressions can be simplified. Introducing the
scalar-to-tensor ratio $r$ and the spectral index $n_s$,
we find
\begin{align}
	\Delta \gNL & = \frac{25}{1152} r (n_s - 1 + 2\epsilon_\ast) \ll 1 \\
	\Delta \tauNL & = -\frac{3}{35} r \fNL ,
\end{align}
where $\fNL$ is the sourced local-mode contribution to the three-point
function.
The $\gNL$ contribution is clearly negligible.
The $\tauNL$ contribution is negligible provided $|r\fNL| \lesssim 1$.
Taking the bound on $r$ to be roughly $r \lesssim 0.1$, this term
can be observationally relevant only if $|\fNL| \gtrsim 50$.
This is already on the verge of being ruled out by experiment, so
the $\tauNL$ contribution is likely to be no more than
$\Or(1)$ in most acceptable
models. It could perhaps be kept if very accurate estimates are required.

Finally, the equilateral-type term is
\begin{equation}
	\begin{split}
	T_\zeta \supseteq \frac{H_\ast^6}{8 k_1^3 k_2^3 k_3^3 k_{14}^3}
	\Bigg(
		&
		N_i N_{ij} \frac{\dot{\varphi}_j}{4H_\ast}
		(N_k N_k)
		\left(
			- \frac{8 k_2^2 k_3^2}{k_t}
			- k_{14} (k_2^2 + k_3^2)
		\right)
	\\ & \mbox{}
		- \frac{N_i N_{ij} N_j}{4}
		\left(
			- 8 \frac{k_{14}^2}{k_t}(k_2^2 + k_3^2) - k_{14}^2 (k_2 + k_3)
			- k_2 k_3 (k_2 + k_3)
		\right)
	\Bigg)
	\\ & \quad \mbox{}
	+ [ \vect{k}_1 \rightarrow \{ \vect{k}_2, \vect{k}_3 \}]
	+ (\text{cyclic $\vect{k}_4 \rightarrow \{
		\vect{k}_1, \vect{k}_2, \vect{k}_3 \}$})
	\end{split}
\end{equation}
The coefficients of these contributions are related to those of
Eqs.~\eqref{eq:effective-g} and~\eqref{eq:effective-tau},
and will therefore not typically be large.

\bibliographystyle{JHEPmodplain}
\footnotesize
\bibliography{paper}

\end{document}